\documentclass[sigconf]{acmart}
\AtBeginDocument{%
  }

\copyrightyear{2026}
\acmYear{2026}
\setcopyright{cc}
\setcctype{by}
\acmConference[HPDC '26]{The 35th International Symposium on High-Performance Parallel and Distributed Computing}{July 13--16, 2026}{Cleveland, OH, USA}
\acmBooktitle{The 35th International Symposium on High-Performance Parallel and Distributed Computing (HPDC '26), July 13--16, 2026, Cleveland, OH, USA}
\acmDOI{10.1145/3806645.3807583}
\acmISBN{979-8-4007-2640-8/2026/07}

\usepackage{booktabs}   

\usepackage{enumitem}
\usepackage{multirow}
\usepackage[ruled,vlined,linesnumbered]{algorithm2e} 

\usepackage{stfloats}

\usepackage{subcaption}
\usepackage{pifont}

\begin{document}

\title[QiankunNet-cuSCI]{A Fully GPU-Accelerated Framework for High-Performance Configuration Interaction Selection with Neural Network Quantum States}

\settopmatter{authorsperrow=4}

\author{Daran Sun}
\authornote{Daran Sun, Bowen Kan, Haoquan Long contributed equally to this work.}
\affiliation{%
\institution{Institute of Computing Technology, Chinese Academy of Sciences}
\city{Beijing}
\country{China}
}
\affiliation{%
\institution{University of Chinese Academy of Sciences}
\city{Beijing}
\country{China}
}
\email{sundaran24s@ict.ac.cn}

\author{Bowen Kan}
\authornotemark[1]
\affiliation{%
\institution{Institute of Computing Technology, Chinese Academy of Sciences}
\city{Beijing}
\country{China}
}
\affiliation{%
\institution{University of Chinese Academy of Sciences}
\city{Beijing}
\country{China}
}
\email{kanbowen17@163.com}

\author{Haoquan Long}
\authornotemark[1]
\affiliation{%
\institution{Institute of Computing Technology, Chinese Academy of Sciences}
\city{Beijing}
\country{China}
}
\affiliation{%
\institution{University of Chinese Academy of Sciences}
\city{Beijing}
\country{China}
}
\email{longhaoquan25@mails.ucas.ac.cn}

\author{Hairui Zhao}
\affiliation{
\institution{Institute of Computing Technology, Chinese Academy of Sciences}
\city{Beijing}
\country{China}
}
\email{zhaohairui@ict.ac.cn}

\author{Haoxu Li}
\affiliation{%
\institution{Institute of Computing Technology, Chinese Academy of Sciences}
\city{Beijing}
\country{China}
}
\email{lihaoxu21@mails.ucas.ac.cn}

\author{Yicheng Liu}
\affiliation{%
\institution{Institute of Computing Technology, Chinese Academy of Sciences}
\city{Beijing}
\country{China}
}
\email{liuyicheng25@mails.ucas.ac.cn}

\author{Pengyu Zhou}
\affiliation{%
\institution{Institute of Computing Technology, Chinese Academy of Sciences}
\city{Beijing}
\country{China}
}
\email{zhoupengyu19@mails.ucas.ac.cn}

\author{Ankang Feng}
\affiliation{%
\institution{University of Science and Technology of China}
\city{Hefei}
\country{China}
}
\email{fak0615@mails.ustc.edu.cn}

\author{Wenjing Huang}
\affiliation{%
\institution{Institute of Computing Technology, Chinese Academy of Sciences}
\city{Beijing}
\country{China}}
\email{huangwenjing23@mails.ucas.ac.cn}

\author{Yida Gu}
\affiliation{%
\institution{Institute of Computing Technology, Chinese Academy of Sciences}
\city{Beijing}
\country{China}}
\email{guyida@ict.ac.cn}

\author{Zhenyu Li}
\affiliation{%
\institution{University of Science and Technology of China}
\city{Hefei}
\country{China}
}
\email{zyli@ustc.edu.cn}

\author{Honghui Shang}
\authornote{Corresponding authors.}
\affiliation{%
\institution{University of Science and Technology of China}
\city{Hefei}
\country{China}
}
\email{	shh@ustc.edu.cn}

\author{Yunquan Zhang}
\affiliation{%
\institution{Institute of Computing Technology, Chinese Academy of Sciences}
\city{Beijing}
\country{China}}
\email{zyq@ict.ac.cn}

\author{Dingwen Tao}
\affiliation{%
\institution{Institute of Computing Technology, Chinese Academy of Sciences}
\city{Beijing}
\country{China}}
\email{taodingwen@ict.ac.cn}

\author{Ninghui Sun}
\affiliation{%
\institution{Institute of Computing Technology, Chinese Academy of Sciences}
\city{Beijing}
\country{China}}
\email{snh@ict.ac.cn}

\author{Guangming Tan}
\authornotemark[2]
\affiliation{%
\institution{Institute of Computing Technology, Chinese Academy of Sciences}
\city{Beijing}
\country{China}}
\email{tgm@ict.ac.cn}

\renewcommand{\shortauthors}{Sun, Kan, and Long et al.}

\captionsetup[figure]{font=small}
\captionsetup[table]{font=small}

\setlength\abovecaptionskip{3pt}
\setlength{\abovedisplayskip}{3pt}
\setlength{\belowdisplayskip}{3pt}
\setlength{\abovedisplayshortskip}{3pt}
\setlength{\belowdisplayshortskip}{3pt}
\setlength\textfloatsep{3pt}

\begin{abstract}
AI-driven methods have demonstrated considerable success in tackling the central challenge of accurately solving the Schrödinger equation for complex many-body systems. Among neural network quantum state (NNQS) approaches, the NNQS-SCI (Selected Configuration Interaction) method stands out as a state-of-the-art technique, recognized for its high accuracy and scalability. However, its application to larger systems is severely constrained by a hybrid CPU-GPU architecture. Specifically, centralized CPU-based global de-duplication creates a severe scalability barrier due to communication bottlenecks, while host-resident coupled-configuration generation induces prohibitive computational overheads.
We introduce QiankunNet-cuSCI, a fully GPU-accelerated SCI framework designed to overcome these bottlenecks. It first integrates a distributed, load-balanced global de-duplication algorithm to minimize redundancy and communication overhead at scale. To address compute limitations, it employs specialized, fine-grained CUDA kernels for exact coupled configuration generation. Finally, to break the single-GPU memory barrier exposed by this full acceleration, it incorporates a GPU memory-centric runtime featuring GPU-side pooling, streaming mini-batches, and overlapped offloading. This design enables much larger configuration spaces and shifts the bottleneck from host-side limitations back to on-device inference. Our evaluation demonstrates that our work fundamentally expands the scale of solvable problems. On an NVIDIA A100 cluster with 64 GPUs, our work achieves up to 2.32$\times$ end-to-end speedup over the highly-optimized NNQS-SCI baseline while preserving the same chemical accuracy. Furthermore, it demonstrates excellent distributed performance, maintaining over 90\% parallel efficiency in strong scaling tests.

\end{abstract}


\maketitle

\section{Introduction}
Electronic structure calculation based on quantum mechanics is an elementary tool for predicting the chemical and physical properties of matter. From a computational perspective, many problems related to material structures and physical
properties can ultimately be reduced to the numerical solution of the Schrödinger equation~\cite{whitfield2013computational}.
In practice, however, this problem has long been constrained by exponential computational
complexity: the state space of many-body quantum systems grows rapidly with system size, rendering
exact methods fundamentally unscalable~\cite{OGorman2022Intractability}. Thus, full configuration interaction (FCI)~\cite{Olsen1990OneBillionFCI}, which enumerates all possible electronic configurations, becomes computationally intractable and is thus limited to small systems even with the best supercomputers~\cite{o2022intractability}.

The advent of neural-network quantum states (NNQS)~\cite{schmitt2020quantum,medvidovic2024neural} has transformed computational quantum chemistry by harnessing artificial neural networks to variationally encode many-body wave functions. 
Carleo and Troyer demonstrated that restricted Boltzmann machines (RBMs) can act as universal approximators of quantum states on lattice spin systems~\cite{carleo2017solving}. This approach offers strong representational capacity together with polynomial-time computational scaling. Since then, NNQS has been successfully applied to a wide range of multi-spin and fermionic systems,
establishing the feasibility and potential of AI-based ansatze~\cite{choo2020fermionic,medvidovic2024neural}.

Building on this foundation, the Transformer architecture—owing to its strong representational
capability across many domains—has been incorporated into NNQS, giving rise to the
NNQS-Transformer paradigm~\cite{qknet2023,ma2024quantum}.
Transformers~\cite{vaswani2017attention} are particularly effective at modeling long-range dependencies and complex
high-dimensional correlations, making them a powerful tool for enhancing ansatz expressiveness~\cite{Sharir2020DeepAutoregressive}. 
Prior NNQS-Transformer frameworks such as QiankunNet~\cite{qknet2023} reduce computational overhead through parallel batch sampling and distributed energy evaluation. Nevertheless, these approaches remain heavily dependent on probabilistic sampling, which can lead to uncontrolled growth of the sample space and introduce systematic approximation errors, ultimately limiting both accuracy and scalability at large scales~\cite{chen2020variational,chen2024empowering}.

To further mitigate sampling-induced errors, subsequent studies integrated NNQS with Selected Configuration Interaction (SCI), yielding NNQS-SCI. QiankunNet-SCI~\cite{Kan2025NNQS} stands out as a state-of-the-art NNQS-SCI technique with superior accuracy and robust scalability; however, its application to larger systems is severely constrained by non-AI workflow elements.

As observed in broader AI-for-Science applications, a critical bottleneck emerges: while AI inference scales efficiently through data parallelism, associated non-AI components---like complex logic and data management---do not, creating scaling imbalances that hinder overall performance~\cite{junaid2025artificial,jia2020pushing,stevens2020ai}.

NNQS-SCI illustrates this clearly, as its scalability is heavily restricted by host-side (CPU) limitations. Global deduplication requires maintaining a single-node index that often reaches hundreds of gigabytes for large molecular systems. This centralized process causes significant GPU underutilization, accounting for 30\% of the total runtime in a 64-GPU configuration, thereby making the master node's CPU memory and processing power the primary bottleneck.

Additionally, coupled-state generation exhibits insufficient scaling efficiency; as GPU counts increase, non-AI runtime escalates sharply from under 10\% to over 50\%. This architectural imbalance stems from executing selected configuration interactions on the CPU while Transformer-based inference scales on the GPU, an issue further exacerbated by frequent host-device data transfers and the CPU-GPU performance gap.

These observations necessitate a clear system-level objective: executing the entire SCI workflow on GPUs to eliminate host-side involvement and transfer overheads. This approach relies on two key insights: (1) distributed deduplication can shift the memory bottleneck from a single CPU to distributed GPUs, and (2) coupling calculations involve uniform operations highly amenable to GPU parallelization. Under this fully GPU-resident design, the performance bottleneck shifts: GPU memory capacity becomes the primary constraint, and neural wavefunction inference re-emerges as the dominant runtime component.

To achieve this, three specific challenges must be addressed. First, to support large molecular systems, we must design a scalable, load-balanced distributed GPU deduplication algorithm. Second, as coupled state generation emerges as the subsequent bottleneck, we must implement high-performance GPU-accelerated configuration calculations. Third, maintaining a fully GPU-resident workflow requires fine-grained memory management to mitigate peak usage and navigate complex stage-wise data dependencies.

To address the above issues, this paper presents the NNQS-SCI framework and makes the following contributions:

\begin{itemize}[leftmargin=1.em,topsep=2pt]
    \item \textbf{A Novel Fully GPU-Accelerated NNQS-SCI Framework:} QiankunNet-cuSCI is a comprehensive parallel scheme that migrates the entire NNQS-SCI workflow, including coupled configuration generation and global deduplication. This framework eliminates the CPU-side scalability bottleneck and minimizes the overhead of frequent host-device data transfers.
    
    \item \textbf{High-Performance Kernels and Distributed Deduplication:} We design specialized CUDA kernels for exact coupled computations using fine-grained parallelism. To support large-scale configuration sets, we implement a distributed, sort-based global deduplication algorithm with hierarchical sampling, achieving deterministic load balancing across multi-GPU clusters.
    
    \item \textbf{GPU Memory-Centric Execution Model:} We introduce a memory efficient management mechanism that treats GPU memory as a first-class constraint. By integrating mini-batch streaming and overlapped offloading, our model resolves the peak memory capacity constraints inherent in large-scale SCI simulations.
    
    \item \textbf{Extensive Experimental Evaluation:} We evaluate QiankunNet-cuSCI across a range of molecular systems. The results show up to 2.32× end-to-end speedup over SOTA hybrid implementations,while improving scalability and shifting the performance bottleneck to GPU inference, thereby establishing a system-level foundation for tackling larger configuration spaces.
\end{itemize}

\section{Background and Motivation}
\label{sec:background}
This section systematically reviews pivotal methodologies in quantum many-body simulations~\cite{choo2019two,hermann2023ab,lange2024architectures,medvidovic2024neural,schmitt2020quantum,Sharir2020DeepAutoregressive,Zhao2022AI4QM_SC22} and gives necessary background for HPC general audience about an AI-HPC combined algorithm called NNQS-SCI. We also outline the motivation behind QiankunNet-cuSCI's design.

\subsection{Selective CI Framework}

We consider the problem of computing the ground-state energy of a quantum system, which can be formulated as the lowest-eigenvalue problem of a large Hermitian matrix,
$H\Psi = E\Psi.$

Rather than explicitly solving this eigenvalue equation, ground-state energy can be obtained by minimizing Rayleigh quotient ~\cite{saad2011numerical}

\begin{equation}
E(\Psi) = \frac{\Psi^\dagger H \Psi}{\Psi^\dagger \Psi}
\end{equation}

which forms the basis of the methods discussed in this work.

Here, \(H\in\mathbb{C}^{d\times d}\) and \(\Psi\in\mathbb{C}^{d}\), where \(d\) is the number of all possible configurations. A configuration is represented by a bitstring of length \(m\) with \(n\) ones (e.g., \(01100010\) m=8 n=3), indicating which \(n\) orbitals are occupied among the \(m\) available orbitals. 
\begin{equation}
\mathcal{H}=\{\text{all bitstrings of length } m \text{ with } n \text{ ones}\}
\end{equation}
\begin{equation}
\notag \implies |\mathcal{H}|=C_m^n,\ d=C_m^n.
\end{equation}
The numerator expands to
\begin{equation}
\Psi^\dagger  H  \Psi
= \sum_{i \in \mathcal{H}} \sum_{j \in \mathcal{H}} \psi_i^* H_{ij} \psi_j
= \sum_{i \in \mathcal{H}} \psi_i^*  \sum_{j \in \mathcal{H}}H_{ij} \psi_j \, .
\end{equation}
This corresponds to the full configuration interaction (FCI) approach~\cite{knowles1984new}. FCI yields the variationally exact energy and is therefore widely used as a benchmark for assessing approximate methods (often referred to as "FCI-level" accuracy)~\cite{gao2024distributed}. However, empirical studies show that only about \(1\%\) of configurations contribute significantly to the ground-state energy~\cite{schriber2017adaptive}. This redundancy motivates Selective Configuration Interaction (SCI) methods~\cite{dral2020quantum}, which retain only the most important configurations while discarding negligible contributions. We define the SCI space as a subset \(\mathcal{S}\subset \mathcal{H}\).

A matrix element \(H_{ij}\) is non-zero if and only if configurations \(i\) and \(j\) differ in the positions of "1"s by at most two. For a fixed configuration \(i\), define the coupled set
\begin{equation}
C_i=\{\, j \mid H_{ij}\neq 0 \,\}.
\end{equation}
To construct \(C_i\), we record the indices of all "1"s in the bitstring of \(i\) and enumerate all configurations obtained by moving one or two of these occupied positions. The resulting configurations are exactly the elements of \(C_i\).

\begin{equation}
 \sum_{i \in \mathcal{H}} \psi_i^* \sum_{j \in \mathcal{H}}  H_{ij} \psi_j\approx
 \sum_{i \in \mathcal{S}} \psi_i^* \sum_{j \in C_i}  H_{ij} \psi_j .
\end{equation}

This is how the general Selective CI method works. The advent of neural-network quantum states (NNQS) has transformed computational quantum chemistry by harness-ing artificial neural networks to variationally encode many-body wave functions. In NNQS-SCI method~\cite{Kan2025NNQS}, an AI model (NNQS-transformer) was used to represent $\psi_i$, It takes the bitstring of configuration $i$ as input, and gives out a complex number $\psi_i$.

\begin{figure}[b]
    \centering
    \includegraphics[width=.98\columnwidth]{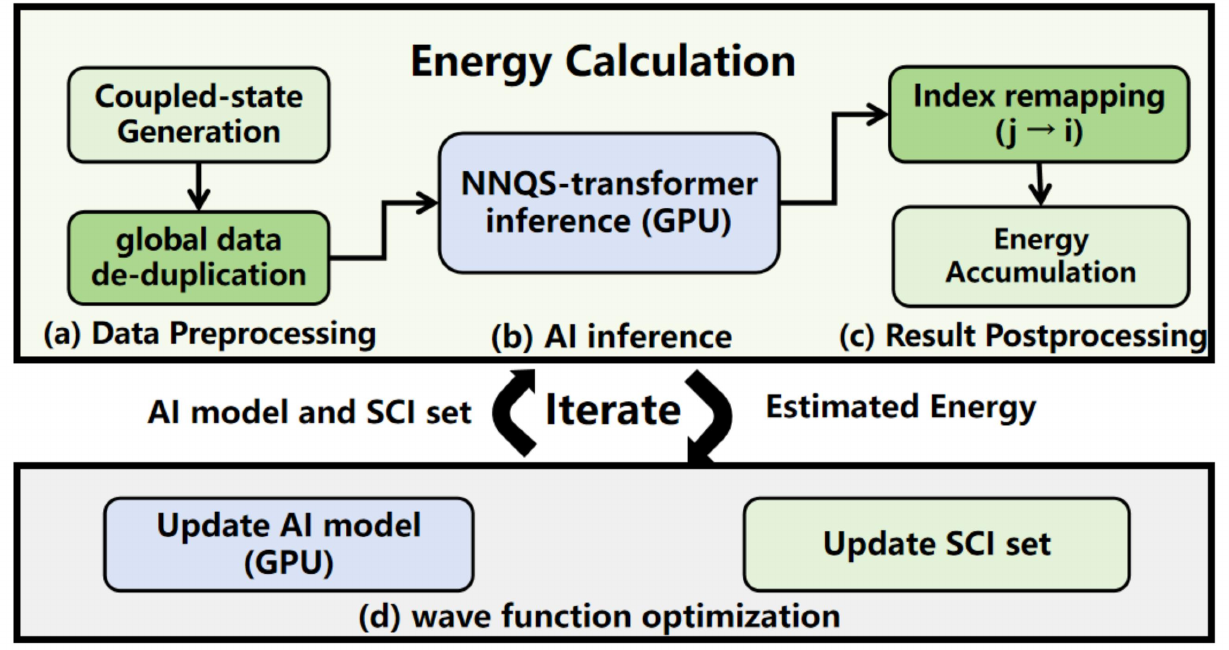}
    \Description{A system workflow diagram of the iterative NNQS-SCI framework, divided into a top panel for Energy Calculation and a bottom panel for wave function optimization. An "Iterate" loop connects the two panels, passing Estimated Energy down and the updated AI model and SCI set up. 

    The top panel shows a three-step pipeline: (a) Data Preprocessing, where coupled-state generation feeds into a CPU-side global data de-duplication block; (b) AI inference, where the unique configurations are processed by the NNQS-transformer on the GPU; and (c) Result Postprocessing, where the output undergoes CPU-side index remapping and energy accumulation. 

    The bottom panel, labeled (d) wave function optimization, contains two separate blocks representing the parameter updates: updating the AI model on the GPU, and updating the SCI set on the CPU.}
    \caption{NNQS-SCI workflow. Green blocks denote CPU-side preprocessing and postprocessing, while NNQS inference and model updates are performed on the GPU.}
    \label{fig:background}
\end{figure}

\subsection{NNQS-SCI workflow}
\label{sub:CI}

NNQS-SCI~\cite{Kan2025NNQS} is an iterative framework that optimizes a variational wavefunction $\Psi$
and estimates its energy $E(\Psi)$ by alternating between (i) energy evaluation and (ii) wavefunction
optimization, the workflow of NNQS-SCI is illustrated in Figure \ref{fig:background}.

\textbf{Energy evaluation.}
Given a selected configuration set $S$, NNQS-SCI enumerates all valid single and double electron
moves for each $i \in S$ to generate its coupled set $C_i$ (Figure \ref{fig:background}(a)).
Materializing all $C_i$ can incur a peak memory footprint of
$O(|S|\cdot m^2 n^2)$ due to the large number of candidate excitations.
To reduce redundant neural inference, NNQS-SCI performs a \emph{global de-duplication} across
$\{C_i\}$ to form a unique configuration set (indexed by $j$) while storing a reverse index mapping
from each unique $j$ back to its originating $i$.
The NNQS-transformer evaluates amplitudes $\psi$ only on unique set (Figure \ref{fig:background}(b)),
followed by index remapping to recover per-$i$ contributions and compute local and global energies
(Figure \ref{fig:background}(c)).

\textbf{Wavefunction optimization.}
The NNQS-transformer parameters are updated by backpropagation using estimated energy.
Meanwhile, NNQS-SCI expands the variational space by selecting a subset of important configurations
from the newly generated candidates (e.g., top-$K$ ranked by inferred amplitudes $\psi$) and merging them
into $S$ for the next iteration. The two phases repeat until convergence.

While conceptually simple, this workflow involves frequent generation, de-duplication, 
and index remapping of large configuration sets across iterations, which become major 
performance bottlenecks in distributed settings.

\begin{figure*}[t]
    \centering 
    \includegraphics[width=.95\linewidth]{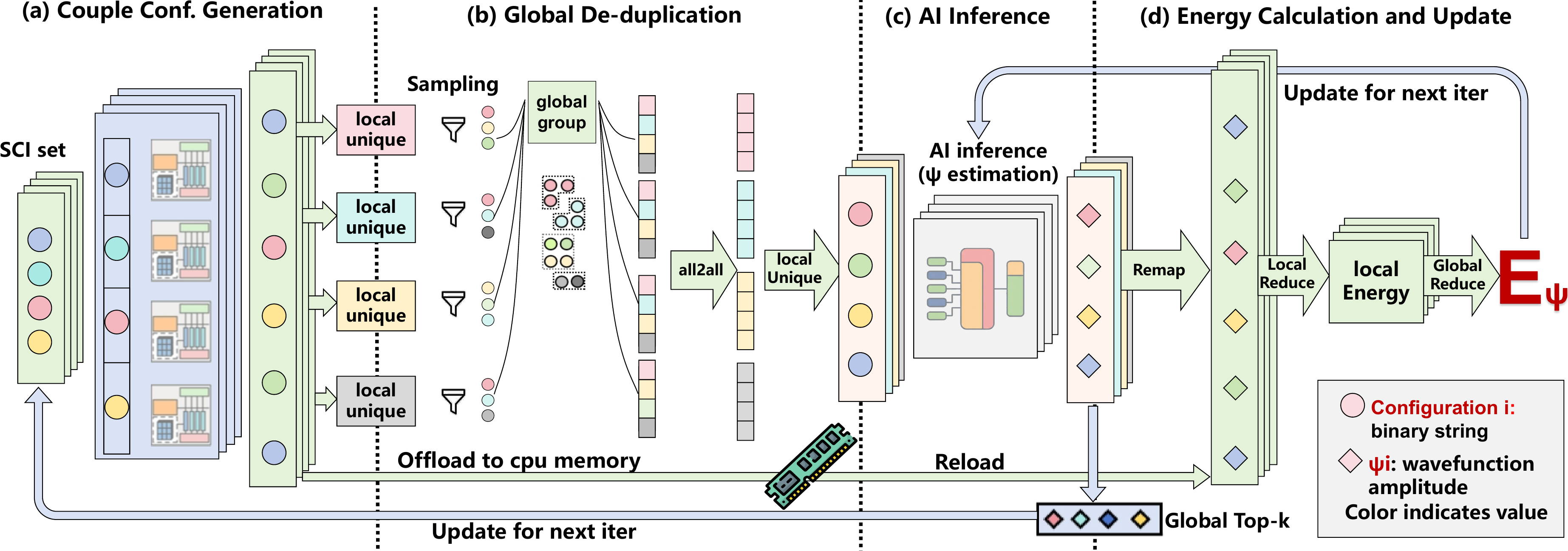}
    \Description{A detailed end-to-end system pipeline diagram divided into four horizontal stages from left to right, with a feedback loop at the bottom. A legend indicates that circles represent binary string configurations and diamonds represent wavefunction amplitudes. 

    Stage (a), Couple Configuration Generation: An initial SCI set of configurations (circles) is processed on the GPU to generate coupled candidates, which then pass through local unique filters. 

    Stage (b), Global De-duplication: The locally unique candidates undergo sampling, global grouping, and an all-to-all data exchange to produce a final set of globally unique configurations. Below stages (a) and (b), an arrow indicates intermediate data being offloaded to CPU memory and later reloaded. 

    Stage (c), AI Inference: The unique configurations (circles) are fed into an AI inference block containing a neural network transformer, which outputs wavefunction amplitudes (diamonds). A bottom path extracts the Global Top-k amplitudes and feeds them back to update the SCI set. 

    Stage (d), Energy Calculation and Update: The amplitudes are remapped to their original configurations, followed by local reduction, local energy calculation, and global reduction to produce the final estimated energy. A top feedback loop labeled "Update for next iter" returns to the start of the pipeline, completing the iterative process.}
    \caption{End-to-end GPU-accelerated QiankunNet-cuSCI pipeline with overlapped execution. The workflow consists of (a) GPU-based coupled configuration generation with local de-duplication, (b) global de-duplication and grouping across devices, (c) batched neural-network inference for wavefunction amplitude estimation, and (d) exact local-energy evaluation followed by reduction and parameter update. Color indicates configuration values and wavefunction amplitudes.}
    \vspace{-2mm}
    \label{fig:overview}
\end{figure*}

\subsection{Motivation}
\label{motivation}
Our analysis of the SOTA NNQS-SCI framework reveals three critical bottlenecks that prevent it from scaling to massive chemical systems, we identify these critical bottlenecks as follows:

\textbf{The Scalability Barrier: Centralized CPU De-duplication.}
As the number of GPUs increases, the inter-node redundancy of generated configurations grows rapidly. The baseline approach relies on a centralized CPU reduction, where a single root node gathers all data. This creates a severe communication bottleneck ($O(N)$ traffic) and memory pressure, causing parallel efficiency to drop significantly at scale (e.g., to 66\% on 64 GPUs). We identify that quantum configuration data, being dense integers, is naturally suited for sorting rather than hashing. This motivates our first contribution: a \textbf{distributed, sort-based de-duplication algorithm} that ensures load balance and minimizes communication overhead.

\textbf{The Compute Bottleneck: CPU-Bound Generation.}
With the de-duplication bottleneck removed, the computational cost shifts to the generation of coupled configurations. In the baseline, this step is performed on CPUs. Our profiling shows that as the system scales, this CPU-bound generation time begins to exceed the GPU inference time, violating the principle of accelerator-centric computing. This imbalance leads to underutilized GPUs and limits the attainable throughput even when sufficient accelerator resources are available. Recognizing that coupled configurations generation involves massive, independent bitwise operations that are inherently amenable to fine-grained GPU parallelism, we propose to put this task on a custom GPU kernel, transforming it from a serial CPU bottleneck into a high-throughput GPU workload.

\textbf{The Capacity Constraint: GPU Memory Wall.} Migrating de-duplication and generation to the GPU creates a fully accelerated pipeline but exposes the limited capacity of High-Bandwidth Memory (HBM). Large chemical systems (e.g., Cr$_{2}$ 84 qubits) generate intermediate configurations that far exceed the 40GB limit of a single A100 GPU, rendering standard in-core execution strategies impossible. To break "memory wall," we introduce a \textbf{GPU memory-centric execution model}. By treating GPU memory as a cache and implementing mini-batch processing with asynchronous offloading, we decouple the peak memory requirement from total problem size, enabling the simulation of previously intractable systems.

\section{QiankunNet-cuSCI Overview}
\label{sec:QiankunNet-cuSCI_overview}

QiankunNet-cuSCI is a GPU framework that integrates neural-network quantum states with
high-precision selected configuration interaction. It follows the same
\emph{iterate--expand--infer--select--optimize} skeleton as NNQS-SCI~\cite{Kan2025NNQS}, but
\emph{redesigns the system pipeline} to (i) remove CPU-side bottlenecks, (ii) enable scalable,
distributed de-duplication, and (iii) support memory-bounded execution via host staging.
Figure~\ref{fig:overview} summarizes the end-to-end dataflow and the newly introduced components. The main iterative loop executes primarily on the GPU,
which serves as the high-performance execution engine. Large intermediate datasets that exceed
GPU memory capacity are treated as \emph{cold data} and are asynchronously staged to and from
CPU host memory. This GPU memory-centric execution model breaks the single-GPU memory barrier while
keeping data movement off the critical path. The workflow proceeds through three stages,
corresponding to panels (a--b), (c), and (d) in Figure~\ref{fig:overview}.

\textbf{Stage 1: Massively Parallel Generation and Global Deduplication.}
Each iteration begins with a set of source configurations from the current SCI space, initialized
from the Hartree--Fock reference. This stage, shown in Figure~\ref{fig:overview}(a--b), constitutes
the core system contribution of QiankunNet-cuSCI and addresses the dominant scalability bottleneck in
prior NNQS-SCI implementations.

The source configurations are processed by a memory-efficient CUDA kernel that exploits fine-grained
GPU parallelism to generate a large number of coupled candidate configurations. To suppress
redundancy early and reduce downstream overheads, local uniqueness filtering is applied immediately
after generation. The remaining candidates are then processed by a scalable global de-duplication
algorithm (Section~\ref{sec:dedup}), producing a globally unique configuration set.
By eliminating inter-node and inter-batch redundancy at scale, this stage ensures that subsequent
inference is performed only on necessary data, significantly reducing GPU memory pressure and compute
cost. Intermediate configuration data are staged to host memory when required, enabling
memory-bounded execution without stalling GPU computation.

\textbf{Stage 2: Batched Inference and Hierarchical Selection.}
In Stage~2, the globally unique configurations are streamed back to the GPU in memory-aware batches
for wavefunction inference. A pre-trained NNQS-Transformer model evaluates the complex wavefunction
amplitudes ($\psi$) for each configuration. To identify the most significant configurations for SCI
expansion while controlling GPU memory usage, QiankunNet-cuSCI employs a two-level hierarchical Top-$K$
selection strategy (Figure~\ref{fig:overview}(c)). Local, intra-batch selection is first applied, and
the surviving candidates are incrementally merged into a running global Top-$K$ set, effectively
pruning the candidate space under a bounded memory footprint.

\textbf{Stage 3: Energy Calculation and Network Optimization.}
The final stage performs exact energy evaluation and neural network optimization. Using a stored
original index, the inferred wavefunction amplitudes $\psi$ are remapped to the original, non-unique
configuration space, enabling accurate local energy computation. Reduction operations aggregate
energy contributions, and the resulting loss signal drives network optimization via standard
backpropagation (Figure~\ref{fig:overview}(d)).
Meanwhile, the Top-$K$ configurations selected in Stage~2 are used to construct an updated SCI
space for the next iteration. With both the SCI space and the neural network parameters refined in Stage~3, the algorithm proceeds to the subsequent iteration, forming a closed iterate--expand--infer--select--optimize loop. This process repeats until convergence is reached.

In the following sections, we present the detailed system and algorithmic designs that enable scalable, memory-efficient execution, with particular emphasis on the GPU-based coupled-state generation and global de-duplication in Stage~1.

\section{Design}

To enable scalable and high-throughput Selected Configuration Interaction (SCI) simulations on GPU clusters, we propose a holistic framework that systematically addresses the bottlenecks of computation, communication, and GPU memory capacity. Our methodology is organized into three synergistic components. 

First, to eliminate the communication overhead caused by data redundancy across nodes, Section \ref{sec:dedup} presents a \textit{distributed global de-duplication algorithm}. By leveraging sort-based regular sampling, this approach ensures deterministic load balancing and strictly coalesced memory access. 
Second, in Section \ref{sec:coupledstates}, we introduce a \textit{fine-grained GPU kernel architecture} co-designed with a compressed data layout. This design resolves the performance--memory trade-off, allowing for the massive generation of coupled configurations with minimal GPU memory footprint. 
Finally, Section \ref{sec:memory_model} defines a \textit{GPU memory-centric execution paradigm}. Through dependency analysis and asynchronous mini-batch pipelining, this model breaks the "memory wall," enabling the simulation of large-scale quantum systems that far exceed physical GPU memory limits.

\begin{figure}[t]
    \centering
    \includegraphics[width=0.92\linewidth]{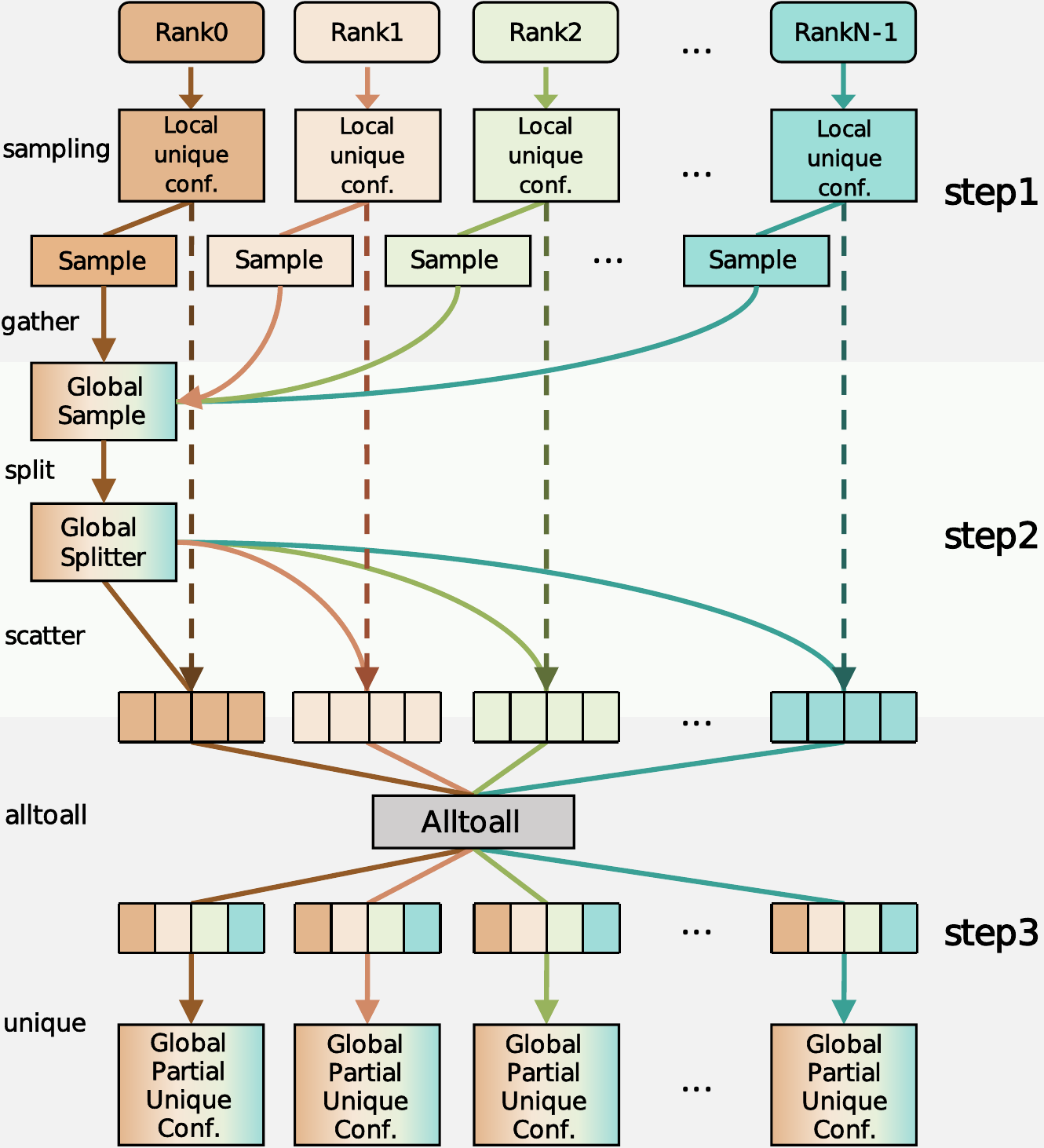}
    \Description{A top-down flowchart illustrating the three-step workflow of a distributed sort-based de-duplication algorithm across multiple parallel processing nodes, labeled Rank 0 to Rank N-1. 

    In step 1, data flows vertically down each rank into "Local unique conf." blocks, from which a "Sample" is extracted. Arrows show these individual samples being gathered into a single "Global Sample" block on the left side.

    In step 2, the "Global Sample" is processed into a "Global Splitter" block. Arrows from the splitter scatter back out to all ranks, indicating how the data is partitioned into local array blocks.

    In step 3, the array blocks from all individual ranks feed into a central "Alltoall" data exchange block. Following this exchange, the reorganized data arrays at each rank undergo a final "unique" operation, culminating in "Global Partial Unique Conf." blocks at the bottom of each respective rank's column.}
    \caption{Workflow of the distributed sort-based de-duplication. By sorting data and sampling at regular intervals, the algorithm ensures that the data volume assigned to each GPU is uniform, preventing load imbalance.}
    \label{fig:unique}
\end{figure}

\subsection{Distributed Global De-duplication via Sort-Based Regular Sampling}
\label{sec:dedup}

Coupled configuration generation is a local operation that produces redundant data across distributed nodes. Eliminating these duplicates is essential to reduce the computational workload of the subsequent neural network inference stage. The baseline framework adopts a \textit{Centralized CPU-Based} approach: nodes gather all local configurations to a single root node, which performs de-duplication using CPU memory before scattering results back, effectively creating a serialization point that leaves other GPUs idle. While simple, this architecture poses fundamental scalability barriers for large-scale quantum simulations, stemming from two critical limitations:  

\textbf{The Memory Wall:} The root node must hold the global dataset in host RAM, limiting the simulation scale to the capacity of a single server and making GPU-resident execution impossible.

\textbf{Communication Bottleneck:} The many-to-one gather operation saturates the root node's bandwidth, creating a serialization point that leaves the massive compute power of GPU cluster idle.

To overcome these limitations, we propose a fully distributed, GPU-resident, de-duplication algorithm that guarantees load balancing and minimizes communication. 

\subsubsection{\textbf{Sort-Based Regular Sampling De-duplication}}
\label{sub:algorithm}

Our approach leverages a deterministic \textbf{Sort-Based Regular Sampling} strategy. By enforcing a global sorted order, we partition the workload based on data volume (rank indices) rather than data values. The algorithm executes in 3 synchronized phases (Figure \ref{fig:unique}):

\textbf{Step 1: Local Preparation and Sample Aggregation.}
Each GPU $i$ first sorts its local configuration buffer $D_i$ using a GPU-optimized Radix Sort. To estimate the global data distribution, we perform regular sampling by selecting $S$ pivots at fixed intervals from the sorted data (specifically at indices $k \times (|D_i| / S)$). These local samples are gathered to the root node. This metadata, with a size of only $P \times S$, provides a lightweight yet representative snapshot of the global dataset, enabling the calculation of optimal global splitters.

\textbf{Step 2: Global Partitioning Strategy.}
The root node sorts the collected samples to determine $P-1$ global splitters at equi-distant intervals. These splitters define the boundaries for $P$ global partitions and are broadcast back to all nodes. Each GPU then performs a binary search on its local sorted array to identify the data range destined for every other rank.

\textbf{Step 3: AlltoAll Exchange and Local Finalization.}
With boundaries established, the cluster executes an \texttt{MPI\_Alltoallv} operation. Rank $i$ sends the data chunk belonging to Partition $j$ directly to Rank $j$. Finally, each GPU performs a local merge and stream compaction. Since the data is globally sorted, duplicates are either adjacent locally or reside at the immediate boundary of neighboring ranks, ensuring complete uniqueness.

Through this structured pipeline, the algorithm guarantees a strictly load-balanced distribution of unique configurations while transforming the irregular redundancy elimination task into efficient, coalesced streaming operations. 

\subsubsection{\textbf{Algorithm Analysis and Design Justification}}
\label{sub:analysis}

Our design choice to utilize sort-based sampling over hash-based partitioning is driven by the specific characteristics of quantum configuration data and the GPU architecture.

\textbf{Efficiency of Radix Sort on GPUs:} 
The quantum configurations are represented as dense arrays of \texttt{uint64} bitmasks. This data format is ideally suited for GPU-optimized Radix Sort, which relies on coalesced memory access and bitwise operations. In contrast, hash-based methods require constructing large index tables with random memory access patterns. Handling collisions via atomic operations (e.g., \texttt{atomicCAS}) on GPUs causes severe thread divergence and serialization, preventing the effective utilization of High-Bandwidth Memory, significantly limiting the overall throughput.

\textbf{Robust Load Balancing via Regular Sampling:} 
Achieving load balancing is critical for distributed performance. Hash partitioning is inherently sensitive to data skew; high-frequency configurations ("heavy hitters") map to the same bucket, creating hot spots. Mitigating this usually requires complex heuristics like "salting" (adding random prefixes), which introduces computational and communication overhead. 
In contrast, our \textbf{Regular Sampling} strategy adaptively captures the global data distribution at a negligible cost. By partitioning based on the sorted order and sampled pivots, the algorithm guarantees that each GPU receives an approximately equal slice of the workload, regardless of the underlying data distribution, effectively eliminating the straggler effect and maximizing resource utilization across the cluster.

\textbf{Minimization of Communication:}
The sort-based layout enables \textbf{Localized Resolution}. The vast majority of duplicates are eliminated locally within a GPU's partition without ever traversing the network. Inter-node redundancy checks are restricted to $O(P)$ boundary elements. This contrasts sharply with hash-based schemes (or the centralized baseline), where communication volume often scales linearly with the total number of duplicates ($O(N)$), resulting in significant bandwidth savings for our approach.

\subsection{GPU Kernel for Coupled Configurations Computation}
\label{sec:coupledstates}

With inter-node scalability resolved by distributed de-duplication, the bottleneck shifts to the CPU-resident coupled-configuration generation, which starves the downstream GPU inference engine due to limited parallelism and PCIe overheads. To address this, we migrate this "massive-generation, sparse-selection" workload entirely to the GPU. However, this introduces a critical performance–memory trade-off: while densely padded structures maximize throughput, they incur prohibitive memory footprints. Consequently, we design a kernel prioritized for strict memory efficiency, co-optimizing data layout and parallel decomposition to achieve substantial acceleration with a minimized footprint, even at the cost of non-coalesced access.

\begin{figure}[t!]
    \centering
    \includegraphics[width=1\linewidth]{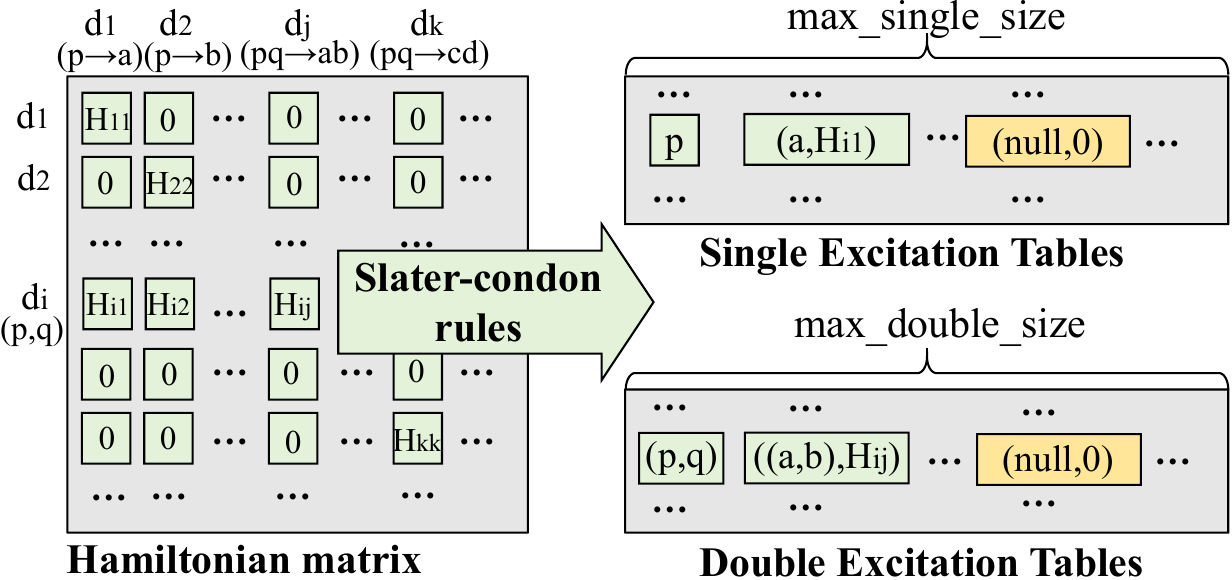}
    \Description{A diagram illustrating the transformation of a sparse Hamiltonian matrix into two compact lookup tables using Slater-Condon rules. 

    On the left, a large grid represents the Hamiltonian matrix, showing source configurations d_i on the vertical axis and target configurations d_1, d_2, d_j, d_k on the horizontal axis. Most elements are zero, with only specific non-zero elements like H_i1 (representing a single excitation) and H_ij (representing a double excitation) highlighted.

    A large arrow labeled "Slater-Condon rules" points from the matrix to the right side, where the data is reorganized into two horizontal arrays. The top array is the "Single Excitation Tables," showing a source orbital p mapping to a target orbital and matrix element pair (a, H_i1). The bottom array is the "Double Excitation Tables," mapping a source pair (p, q) to a target pair and matrix element ((a, b), H_ij). Both tables are padded with (null, 0) elements to reach a uniform maximum size.}
    \caption{For a given source configurations $d_i$(with electrons in orbitals p,q) only considers its connections to other configurations. $H_{i1}$: single excitation (p $\rightarrow$ a), $H_{ij}$: double excitation (pq $\rightarrow$ ab). Applying Slater-Condon rules to Hamiltonian matrix to get single Excitation Table as $T_{single}$ and double Excitation Table as $T_{double}$.
    }
    \label{fig:slater-condon}
\end{figure}

\subsubsection{\textbf{Data Layout Co-Design for Memory Efficiency}}

\label{sub:data_layout}

Generating coupled configurations is computationally equivalent to probing the sparse, off-diagonal structure of the global Hamiltonian matrix, $\mathbf{H}$. The Slater-Condon rules dictate that non-zero matrix elements $H_{ij}$ only exist if configuration $d_j$ is a single or double electronic excitation from $d_i$. Instead of storing this vast matrix, we pre-process its rules into two highly compressed \textbf{Excitation Tables}: $T_{single}$ and $T_{double}$ (Figure \ref{fig:slater-condon}). These tables store pre-calculated integrals, indexed by source and target orbitals. 

To illustrate the effectiveness of this compression, consider the N2 molecule (14 electrons, 56 orbitals). The full Hamiltonian matrix dimension would be $C(56, 14) \approx 4.3 \times 10^{12}$, requiring exabytes of storage. In contrast, our compressed tables are constructed by padding all excitation lists to a uniform length (max\_single\_size=27, max\_double\_size=354). The total memory footprint for these tables is less than 400 KB, a reduction of over 15 orders of magnitude. This aggressive memory optimization comes with a deliberate cost: it enforces an unavoidable indirect addressing (gather) pattern when accessing the lookup tables, making the kernel memory-bandwidth-bound. We accept this as a principled trade-off to stay within realistic device memory budgets and keep the workflow on GPUs.

\newcommand{\GlobalKw}{\textbf{\_\_global\_\_}}
\newcommand{\SharedKw}{\textbf{\_\_shared\_\_}}
\newcommand{\DeviceKw}{\textbf{\_\_device\_\_}}
\definecolor{myCommentGreen}{RGB}{54,102, 71}
\newcommand{\mygreencomment}[1]{\textcolor{myCommentGreen}{// #1}}
\begin{algorithm}[t]
\footnotesize
\sffamily
\SetAlgoLined
\caption{An Overview of the High-Level Structure for the Coupled-Configuration Generation Kernel}
\label{alg:kernel_main}
\DontPrintSemicolon
\textbf{Function:} \GlobalKw \textsc{coupledKernel}(
    SourceDet* source\_dets, 
    ExcitationTable* T\_double, 
    Result* output\_buffer, 
    int* g\_counter
)
\Begin{
    \SharedKw Result temp\_results[BLOCK\_SIZE]\;
    \SharedKw int valid\_mask[BLOCK\_SIZE]\;
    \mygreencomment{Block-level grid-stride loop over source configurations}
    \For{source\_idx = blockIdx.x; source\_idx < num\_sources; source\_idx += gridDim.x}{
        \mygreencomment{Initialize shared memory for the current source configurations}
        valid\_mask[threadIdx.x] = 0\;
        \textbf{\_\_syncthreads}()\; 
        \mygreencomment{Thread-level loop over a virtual space of excitations}
        \For{virtual\_id = threadIdx.x; virtual\_id < $N_\mathit{double}$; virtual\_id += blockDim.x}{
            \mygreencomment{Focus on double excitations for clarity}
            pair\_idx, target\_idx $\gets$ DecomposeVirtualID(virtual\_id)\;
            source\_orbs(i,j) $\gets$ GetOrbitalPair(source\_dets[source\_idx], pair\_idx)\;
            target\_orbs(a,b) $\gets$ GetOrbitalPairFromTable(T\_double, target\_idx)\;
            H\_element $\gets$ IndexedRead(T\_double, source\_orbs, target\_orbs)\;
            \If{$\lvert$H\_element$\rvert > \epsilon$}{
                new\_det $\gets$ GenerateConfigurations(source\_dets[source\_idx], i, j, a, b)\;
                valid\_mask[threadIdx.x] = 1\;
                temp\_results[threadIdx.x] = \{new\_det, H\_element\}\;
            }
        }
        \textbf{\_\_syncthreads}()\;
        \mygreencomment{Filter and write valid results using a modular device function}
        CompactFunc(temp\_results, valid\_mask, output\_buffer, g\_counter)\;
        \textbf{\_\_syncthreads}()\; 
    }
} 
\end{algorithm}

\subsubsection{\textbf{Fine-Grained Kernel Architecture and Execution Flow}}

\label{sub:kernel_design}
\begin{figure}[t!]
    \centering
    \includegraphics[width=0.96\linewidth]{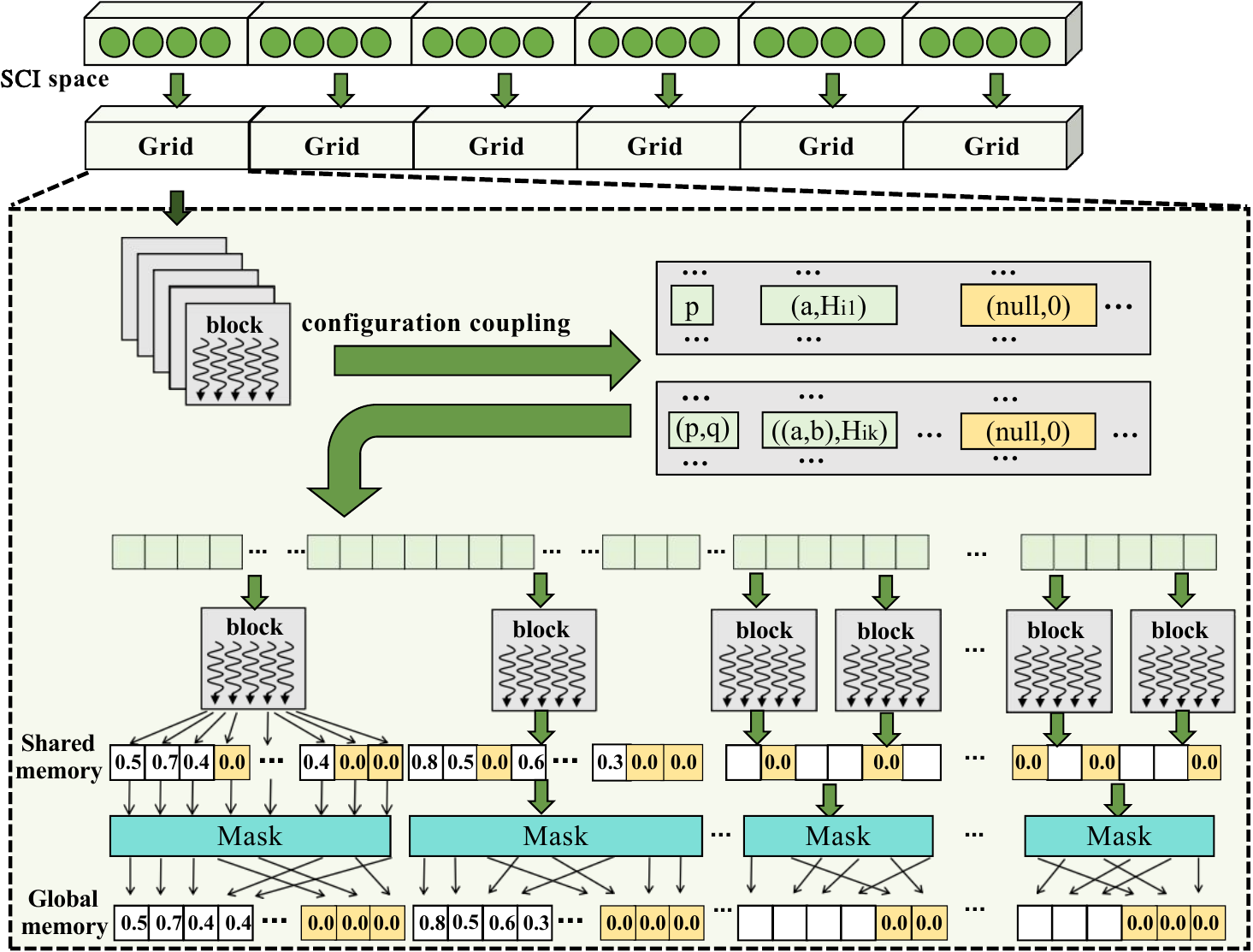}
    \Description{A three-tier hierarchical diagram illustrating the GPU execution flow. The top tier shows an SCI space array divided into a Grid. The middle tier zooms into a Grid element, showing parallel thread blocks using data from excitation tables to perform configuration coupling. The bottom tier illustrates the filtering process: numerical results are stored in shared memory, passed through a "Mask" layer to filter out zero values, and the remaining valid numbers are compacted into a dense array in global memory.}
    \caption{GPU-accelerated SCI algorithm with dual screening criteria (non-zero integrals + adaptive thresholds) enables efficient quantum state generation and rapid many-body Hamiltonian construction via hierarchical grid-block architecture.}
    \label{fig:GPU_couple_state}
\end{figure}

Our kernel design (Algorithm \ref{alg:kernel_main}) transforms the heavy, monolithic CPU task into a fine-grained, two-level parallel workload perfectly suited for the GPU architecture, as depicted in Figure \ref{fig:GPU_couple_state}.

\textbf{Parallel Decomposition.} We employ a "one block per source configuration" mapping. This allows all threads within a block to cooperate on the same source data using low-latency \_\_shared\_\_ memory. Within each block, we assign "one thread per potential excitation". To amortize the significant CUDA kernel launch overhead, a persistent kernel with a two-level loop structure is used. An inner loop iterates over a virtualized space of excitations. We define a total number of virtual threads required to cover all possible single and double excitations for a configurations with $n_{elec}$ electrons:
\begin{align}
    N_{single} &= n_{elec} \times \text{max\_single\_size}. \\
    N_{double} &= \frac{n_{elec} \times (n_{elec} - 1)}{2} \times \text{max\_double\_size}.
\end{align}
Each physical thread in the block processes multiple virtual threads in a loop. As shown in Algorithm \ref{alg:kernel_main} (lines 8-10), each thread decomposes its virtual\_id to determine the specific source orbitals (e.g., pair $(i,j)$) and target orbital index. It performs an indexed read from corresponding excitation table to retrieve the Hamiltonian matrix element, $H_{element}$, and computes new configurations via bitwise operations to minimize latency.

\textbf{Execution Flow and High-Throughput Filtering.}  Each thread first computes its target excitation based on its virtual ID. The thread then performs an indexed read to fetch the corresponding Hamiltonian element $H_{element}$ (line 11). A critical challenge is that only a sparse subset of these generated configurations are valid (i.e., $|H_{element}| > \epsilon$). The configuration itself is represented as a bitmask (uint64\_t array), allowing the creation of a new configurations to be a single, highly efficient bitwise XOR operation (line 13). To filter these efficiently without the serialization penalty of atomicAdd, we implement an integrated parallel stream compaction. After the computation loop, threads with valid results collaboratively perform a fast, block-wide prefix sum on a valid\_mask in shared memory (line 14). This gives each valid result a conflict-free destination index within a temporary shared buffer. Finally, the entire block performs a single, coalesced write to transfer its compacted results to global memory, using a single atomicAdd per block to reserve its destination slab. This design ensures that both the generation and the filtering stages are executed with maximum parallelism and minimal memory contention.

\subsection{A GPU Memory-Centric Execution Paradigm for Scalable Scientific Computing}
\label{sec:memory_model}

With computational and distributed bottlenecks addressed, the final barrier to simulating larger systems is the node-level memory wall. The working set of a single SCI iteration can easily exceed the capacity of high-bandwidth device memory (HBM). To overcome this, we propose a generalizable GPU memory-centric execution paradigm tailored for large-scale AI for Science (AI4S) applications. Our approach systematically addresses the memory constraint through three principles: (1) rigorous data dependency analysis to identify synchronization barriers, (2) mini-batch processing to cap peak memory usage, and (3) asynchronous compute-transfer overlap to hide host-device communication latency.

\subsubsection{\textbf{Dependency Analysis and Stage Partitioning}}

A naive streaming approach is often precluded in scientific computing by global dependencies. We analyzed the intrinsic data flow of the SCI algorithm and identified two unavoidable global synchronization barriers that require a complete view of the global dataset, effectively partitioning the workflow into three distinct computational stages:
\textbf{The Global De-duplication Barrier:} The uniqueness of a configuration cannot be determined until all candidate coupled configurations from all nodes have been generated and aggregated. This forces a synchronization point where the entire candidate set must be materialized (in host memory) before inference can begin.
\textbf{The Selection \& Restoration Barrier:} Global Top-K selection requires the wavefunction amplitudes ($\Psi$) of all unique configurations. Furthermore, mapping these unique $\Psi$ values back to the original coupled set for energy calculation requires a complete reverse index. Based on these barriers, we restructured the monolithic iteration into a three-stage macro-pipeline. This separation allows us to manage memory constraints independently for each stage.

\begin{figure}[t]
    \centering
    \includegraphics[width=0.95\linewidth]{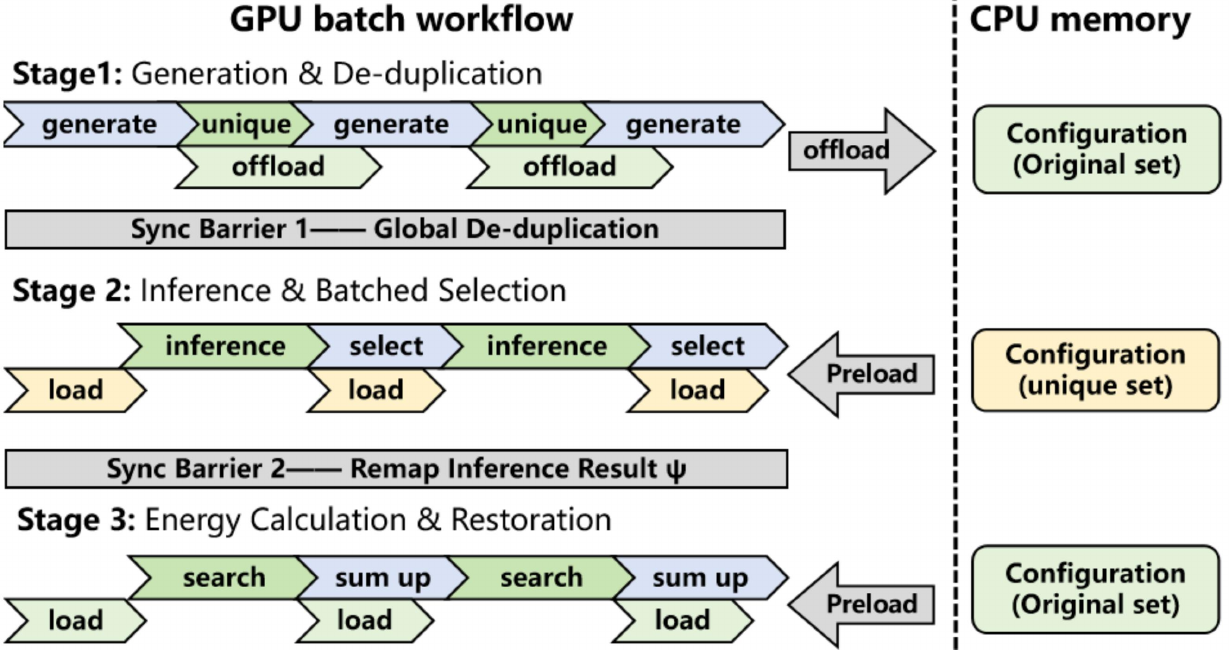}
    \Description{A workflow diagram illustrating a three-stage GPU pipeline interacting with CPU memory to demonstrate overlapped computation and data transfer. The diagram is split into "GPU batch workflow" on the left and "CPU memory" on the right. 

    Stage 1, "Generation and De-duplication", shows a sequence of overlapping chevron arrows representing GPU compute tasks (generate, unique) running in parallel with data transfer tasks (offload). The data is offloaded to the CPU memory and stored as the "Original set". 

    A horizontal bar marks "Sync Barrier 1: Global De-duplication", separating the first two stages.

    Stage 2, "Inference and Batched Selection", shows parallel GPU compute tasks (inference, select) overlapping with "load" tasks. Here, data is preloaded from the CPU memory's "unique set" into the GPU. 

    A second horizontal bar marks "Sync Barrier 2: Remap Inference Result", separating the final stages.

    Stage 3, "Energy Calculation and Restoration", displays parallel GPU compute tasks (search, sum up) overlapping with "load" tasks, as the "Original set" is preloaded back from the CPU memory.}
    \caption{Staged GPU workflow with overlapped host–device data movement. Generation, inference, and energy calculation are executed in GPU batches, while configuration offloading and preloading from CPU memory are overlapped across stages.}
    \label{fig:mem_flow}
\end{figure}

\subsubsection{\textbf{Mini-Batch Processing for Peak Memory Reduction}}

To execute these stages within limited GPU memory, we treat the device memory not as a storage medium but as a high-speed "scratchpad" (cache) for the active working set. We implement a Mini-Batch Execution Model where large datasets are sliced into manageable chunks ($B_{size}$), processed sequentially to enforce memory ceiling.

Instead of processing the entire dataset $D$ of size $N$, we iterate through $N/B_{size}$ batches. For instance, during the inference stage, rather than storing all output logits, we immediately perform a local reduction (e.g., local Top-K update) on the current batch and discard the raw data. This ensures that the peak GPU memory footprint is determined solely by $B_{size}$ and model weights, effectively decoupling it from the total problem scale $N$.

\subsubsection{\textbf{Hiding Latency via Asynchronous Compute-Transfer Overlap}} While offloading "cold" data to host memory solves the capacity issue, it introduces significant latency due to PCIe bandwidth bottlenecks. To mitigate this, we employ a multi-stream pipelining mechanism that overlaps data transfer with computation.

We utilize separate CUDA streams for Host-to-Device (H2D) transfer, Kernel Computation (Compute), and Device-to-Host (D2H) transfer. By employing a double-buffering strategy, we pre-fetch batch $i+1$ and write back batch $i-1$ while the GPU computes batch $i$. This 3-way overlap strategy hides the communication overhead associated with the frequent host-device data movement, ensuring that GPU execution units remain saturated even when processing massive out-of-core datasets.

\subsubsection{\textbf{Implementation: The Three-Stage Pipeline}}

Applying this paradigm, the memory management for each stage is as follows:

\textbf{Stage 1 (Generation \& De-duplication):} As coupled configurations are generated on the GPU in batches, cold data are immediately offloaded to host memory via an asynchronous D2H stream, while only inference-relevant data are retained on the GPU, preventing memory accumulation. Global de-duplication is then performed on the retained device data directly on the GPU.

\textbf{Stage 2 (Inference \& Batched Selection):} Unique configurations are streamed onto the GPU (H2D) in batches for inference. To avoid storing all resulting $\Psi$ values, we utilize a streaming reduction approach: a running, heap-based collection of the best candidates is maintained in GPU memory and progressively refined with each new batch, discarding non-essential amplitudes immediately.

\textbf{Stage 3 (Energy Calculation \& Restoration):} To map $\Psi$ values back, we process the original, non-unique configurations in batches. For each batch, we construct the required reverse index "just-in-time" by searching against the full unique set (streamed from host memory) on GPU. This strategy avoids ever materializing the entire, massive reverse index in GPU memory.

\section{Experimental Evaluation}
\label{sec:evaluation}

In this section, we present a comprehensive evaluation of QiankunNet-cuSCI to demonstrate both its numerical correctness and system-level efficiency. We begin in Section \ref{sec:eval_acc} by validating that our extensive system redesign and GPU-centric optimizations do not compromise the accuracy of the SOTA NNQS method, QiankunNet-SCI~\cite{Kan2025NNQS}.
In Section \ref{sec:eva_per}, we analyze the end-to-end performance gains and bottleneck shifts on a large-scale distributed GPU cluster, followed by a detailed scalability analysis in Section \ref{sec:eva_scal}.
Subsequently, Section \ref{sec:eva_dis_de} evaluates the robustness of our distributed global de-duplication algorithm, with a specific focus on load balancing and throughput across diverse chemical systems.
Finally, Section \ref{sec:eva_mem} demonstrates the effectiveness of GPU memory-centric execution model in overcoming hardware memory constraints for memory-intensive workloads.

\subsection{Evaluation Setup}
\label{sub:setup}
\textbf{Platforms}
The experimental evaluation is conducted on a heterogeneous cluster consisting of up to 16 compute nodes. Each node is equipped with a Kunpeng-920 CPU (128 cores) and 256 GB of host memory. For acceleration, each node features four NVIDIA A100 GPUs (40GB) connected via PCIe. Inter-node communication is established through a high-speed Ethernet network.

On the software side, QiankunNet-cuSCI is implemented using CUDA C++ for coupled configuration calculations, leveraging the CUB and Thrust libraries. The implementation is integrated as a PyTorch C++ extension. The framework utilizes PyTorch to manage the Transformer decoder and torch.distributed for parallelization. Additionally, CuPy is employed to compute local energies efficiently during the wave function optimization process.

The designs of our distributed global de-duplication algorithm and coupled calculation kernel are platform-agnostic and can be implemented on different GPU platforms. In this work, we selected the NVIDIA A100 platform to evaluate these designs.

\textbf{NNQS models and datasets} 
The wave function ansatz in this work is configured as follows: for the amplitude part, we set the embedding dimension to 32, 4 decoder layers, and 4 attention heads. And for the phase part, we use 4-layer MLP with hidden dimensions [512, 512, 512]. We have used AdamW [21] as the gradient descent optimizer with learning rate setting to $3\times10^{-4}$. To comprehensively evaluate the accuracy and scalability of QiankunNet-cuSCI, we selected a diverse set of chemical systems ranging from simple molecules to complex, strongly correlated systems. These benchmarks are categorized into three tiers based on their qubit count:

\begin{itemize}[leftmargin=1.em,topsep=2pt]
    \item \textbf{Small-scale Benchmarks ($\le 30$ qubits):} We utilize $\text{C}_2$ and $\text{N}_2$ under the minimal STO-3G basis set, alongside $\text{LiH}$, $\text{LiF}$, $\text{LiCl}$, and $\text{Li}_2\text{O}$. These systems serve as the primary baseline for verifying numerical correctness. The ground-state reference energies for these molecules are exactly computed using the Full Configuration Interaction (FCI) method via the PySCF library~\cite{sun2018pyscf}.
    
    \item \textbf{Medium-scale Systems ($30 \sim 64$ qubits):} To evaluate performance on intermediate workloads, we select $\text{C}_2\text{H}_4\text{O}$ and $\text{H}_2\text{O}$, as well as $\text{C}_2$ and $\text{N}_2$ employing the more comprehensive cc-pVDZ basis set. The application of this larger basis set significantly expands the configuration space and required qubit count. The reference energies for these systems are similarly obtained from PySCF calculations.
    
    \item \textbf{Large-scale Challenge (84 qubits):} To stress-test the framework's limits on strongly correlated systems, we study the Chromium dimer ($\text{Cr}_2$), which requires 84 qubits. As FCI is computationally intractable at this scale, we instead adopt the highly accurate result from the SOTA QiankunNet-SCI~\cite{Kan2025NNQS} method as the baseline reference ground-state energy.
\end{itemize}

\begin{figure}[b]
    \centering
    \includegraphics[width=0.96\linewidth]{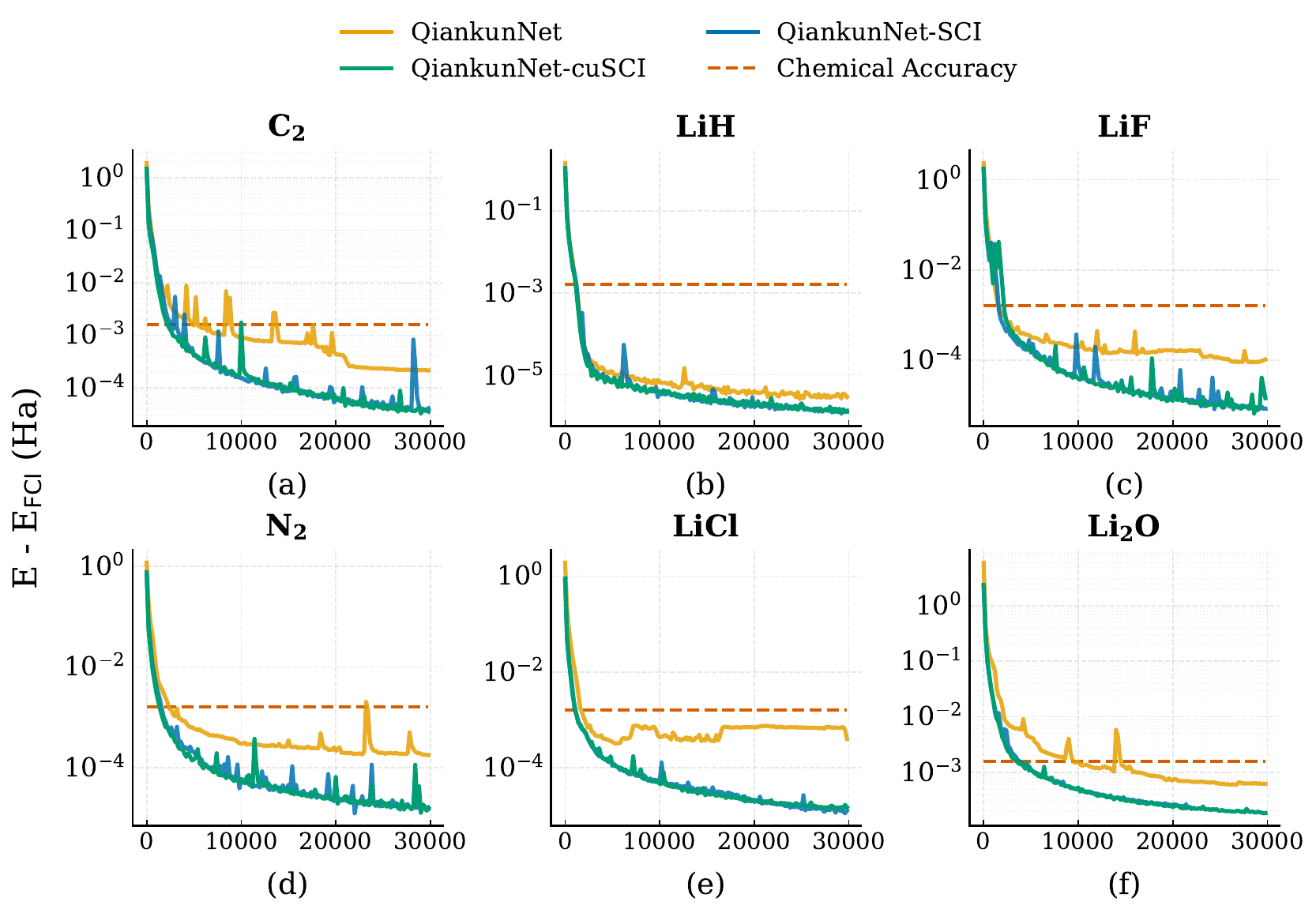}
    \Description{A 2 by 3 grid of line graphs showing the absolute energy errors over 30,000 iteration steps for six molecules: C2, LiH, LiF, N2, LiCl, and Li2O. The y-axis is on a logarithmic scale. In all six graphs, a horizontal dashed line marks the chemical accuracy threshold. The line for the NNQS-Transformer method generally remains higher and noisier. In contrast, the lines for the QiankunNet-SCI and QiankunNet-cuSCI methods overlap almost perfectly, dropping sharply in early iterations and stabilizing well below the chemical accuracy dashed line for all molecules.}
    \caption{Subfigures (a)--(f) present the absolute energy errors (with respect to full configuration interaction, FCI) as functions of iteration steps for  $\text{C}_2$ and  $\text{N}_2$ (sto3g), LiH, LiF, LiCl, Li2O, respectively, calculated using the NNQS-Transformer, QiankunNet-SCI and QiankunNet-cuSCI methods. The red dashed line denotes the chemical accuracy threshold ($0.0016$~hartree).}
    \label{fig:eva_6mols}
\end{figure}

\subsection{Evaluation of Accuracy}
\label{sec:eval_acc}

This section verifies that QiankunNet-cuSCI's fully GPU-accelerated framework maintains the numerical accuracy of the underlying SCI methodology. We benchmark against two distinct baselines: the sampling-based NNQS-Transformer method, QiankunNet, and the SOTA exact NNQS-SCI method, QiankunNet-SCI.

First, we evaluate small-scale systems ($\le 30$ qubits): $\text{C}_2$ and $\text{N}_2$ (STO-3G), $\text{LiH}$, $\text{LiF}$, $\text{LiCl}$, and $\text{Li}_2\text{O}$. As Figure~\ref{fig:eva_6mols} shows, QiankunNet-cuSCI exhibits superior convergence, consistently stabilizing below the chemical accuracy threshold ($1.6 \times 10^{-3}$ Hartree) across all molecules, whereas QiankunNet struggles to reach comparable precision. Furthermore, upon convergence, the ground-state energies computed by QiankunNet-cuSCI perfectly match those of the exact QiankunNet-SCI baseline, confirming that our GPU approach preserves the high fidelity of the original formulation.

\begin{figure}[t]
    \centering
    \includegraphics[width=0.96\linewidth]{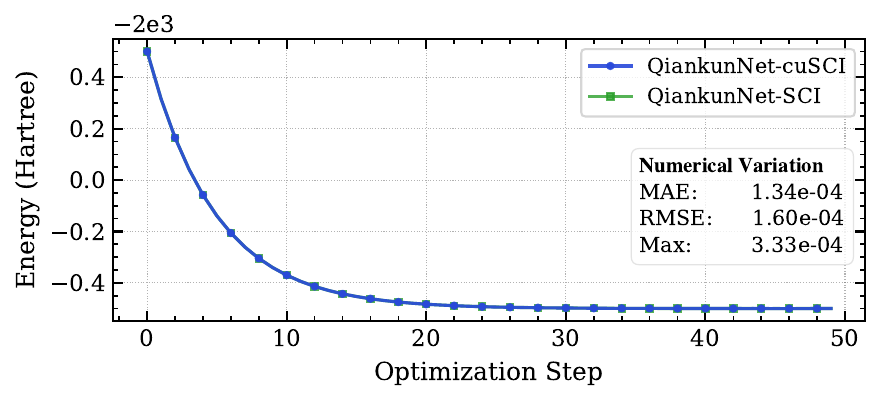}
    \Description{A line graph comparing the step-by-step energy trajectories of the QiankunNet-cuSCI and QiankunNet-SCI methods for the Cr2 system over roughly 110 optimization steps. The energy curves for both methods overlap almost perfectly, showing a rapid initial descent that smoothly levels off. An inset text box quantifies the extremely small numerical variation between the two methods, listing a Mean Absolute Error of 3.91e-04, a Root Mean Square Error of 5.34e-04, and a Maximum error of 2.06e-03.}
    \caption{Step-by-step energy comparison on $\text{Cr}_2$ between QiankunNet-cuSCI and QiankunNet-SCI. The discrepancy is quantified by MAE (mean absolute error), RMSE (root mean square error), and Max (maximum absolute error).}
    \label{fig:compare_energy}
\end{figure}

To assess accuracy at scale, we evaluate the strongly correlated 84-qubit $\text{Cr}_2$ system. Since this exceeds QiankunNet's computational limits, we compare exclusively against QiankunNet-SCI. Figure~\ref{fig:compare_energy} presents the step-by-step energy trajectories. Both methods exhibit nearly identical optimization paths, demonstrating that our work scales to massive systems without accuracy loss.

Note that QiankunNet-cuSCI does not achieve strict bit-to-bit equivalence with QiankunNet-SCI. This deviation stems inherently from hardware execution differences rather than algorithmic flaws. Floating-point arithmetic is non-associative; while QiankunNet-SCI accumulates amplitudes and local energies sequentially on CPUs, QiankunNet-cuSCI employs massively parallel GPU reductions and atomic operations. The non-deterministic summation order of thousands of concurrent threads introduces micro-level rounding differences. However, as quantified in Figure~\ref{fig:compare_energy}, the extremely low deviation (MAE of $3.91 \times 10^{-4}$ Hartree for the 84-qubit $\text{Cr}_2$) and the lack of error accumulation over iterations prove this variance is benign. Thus, our work maintains rigorous SCI accuracy standards while delivering significant performance acceleration.

\subsection{End-to-End Performance and Optimization Breakdown}
\label{sec:eva_per}

To evaluate the practical efficacy of QiankunNet-cuSCI in large-scale high-performance computing scenarios, we conducted distributed experiments on a cluster of 64 NVIDIA A100 GPUs. We employed two representative systems with distinct computational characteristics selected to stress system bottlenecks:
(1) \textbf{Cr$_2$ System:} A heavy-generation workload initialized with 32,000 source configurations, where each generates approximately 30,000 coupled configurations (totaling $\sim 10^9$ candidates).
(2) \textbf{N$_2$ System:} A large-basis workload (cc-pVDZ) initialized with 256,000 source configurations, each generates $\sim 4,100$ configurations.

\begin{figure}[t!]
    \centering
    \includegraphics[width=0.96\linewidth]{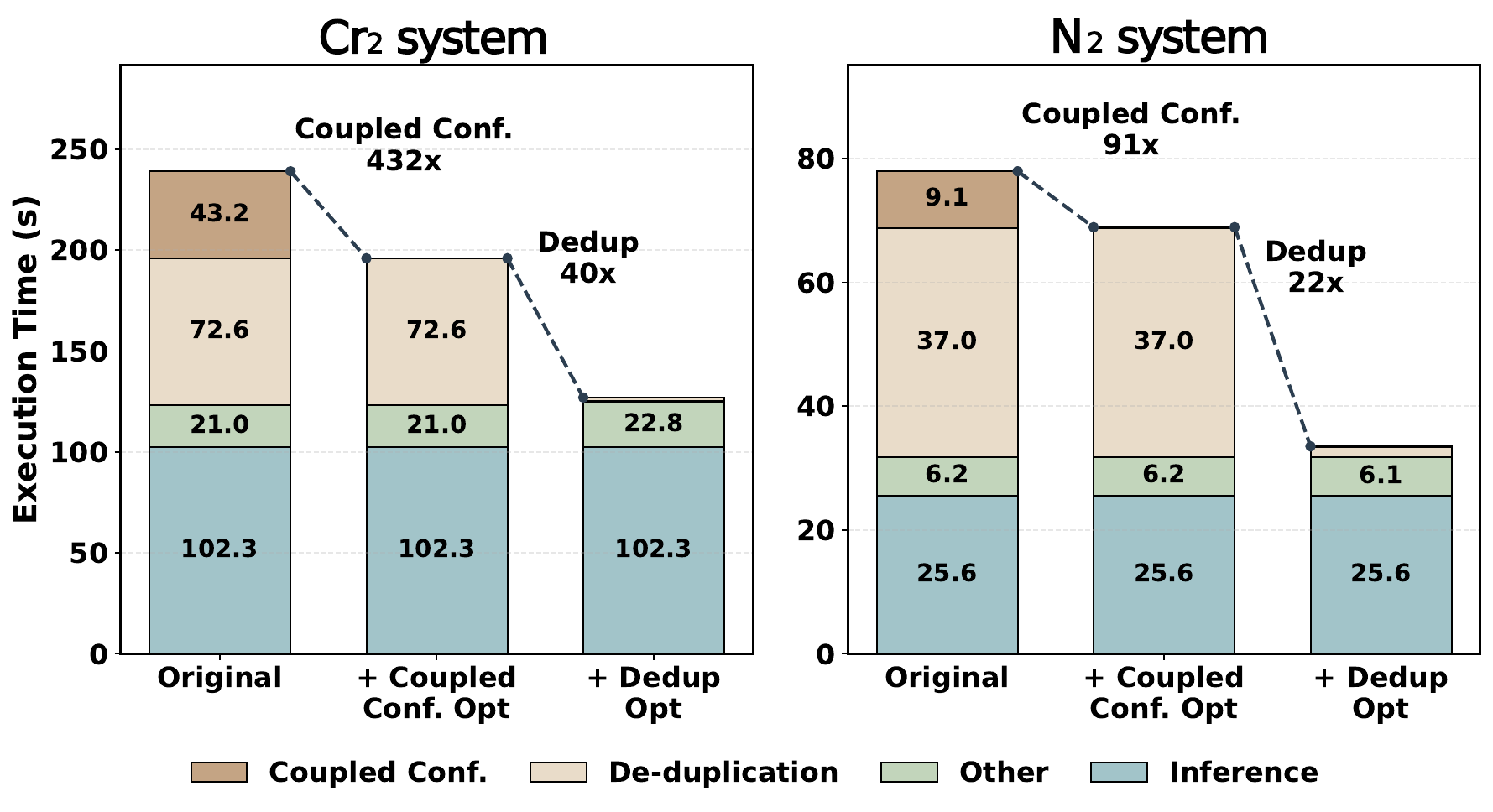}
    \Description{Two side-by-side stacked bar charts comparing the execution time breakdown for the Cr2 and N2 systems. The y-axis shows Execution Time in seconds. Each chart displays three bars representing a progression: Original, plus Coupled Configuration Optimization, and plus De-duplication Optimization. 

    The bars are stacked with four categories from bottom to top: Inference, Other, De-duplication, and Coupled Configuration. 

    In the "Original" bars, the top two segments (De-duplication and Coupled Configuration) consume a large portion of the total height. 

    In the middle bars, dashed lines highlight the topmost Coupled Configuration segment shrinking to a thin sliver, annotated with a 432x reduction for Cr2 and a 91x reduction for N2. 

    In the rightmost bars, dashed lines show the De-duplication segment also shrinking drastically to almost nothing, annotated with a 40x reduction for Cr2 and a 22x reduction for N2. 

    Visually, the final optimized bars are much shorter overall, and the bottom "Inference" segment, which maintains a constant height across all bars, becomes the overwhelmingly dominant portion of the total execution time.}
    \caption{End-to-end execution time breakdown for Cr$_2$ and N$_2$ systems under the 64-GPU distributed environment.}
    \label{fig:end2end}
\end{figure}

We compared our fully GPU-accelerated framework against the SOTA QiankunNet-SCI baseline. To ensure a fair comparison, the neural network inference components in both setups are identical and executed on GPUs. For the baseline's CPU-dependent stages, we utilized a highly parallelized implementation running on 128-core Kunpeng-920 CPUs, representing the CPU performance. Figure \ref{fig:end2end} illustrates the step-by-step execution time breakdown, revealing how each optimization contributes to total speedup.

\textit{1. Baseline Bottlenecks (Original):} 
As shown in the "Original" bars, the baseline suffers significantly from non-AI overheads. For the heavy-generation Cr$_2$ system, the coupled configuration generation takes 43.2 seconds, and the centralized de-duplication consumes 72.6 seconds. Together with other overheads, non-inference tasks account for 57\% of the total runtime (239.1s), exceeding the AI inference time (102.3s). Similarly, for N$_2$, the non-inference stages consume 67\% of the total time (52.3s out of 77.9s), with de-duplication being the primary bottleneck (37.0s) due to the large number of source configurations, limiting further distributed scalability.

\textit{2. Optimization 1: GPU Coupled-Configuration Generation (+Coupled Conf. Opt):} 
By offloading the generation task to our bitwise-optimized CUDA kernel, we observe an order-of-magnitude performance leap.
For Cr$_2$, the generation time drops precipitously from 43.2s to 0.1s, achieving a \textbf{432$\times$ speedup} for this specific kernel.
For N$_2$, the generation time is reduced from 9.1s to 0.1s, a \textbf{91$\times$ speedup}.
This effectively eliminates the generation phase as a bottleneck, as shown in the middle bars of Figure \ref{fig:end2end}.

\textit{3. Optimization 2: Distributed Global De-duplication (+Dedup Opt):} 
The final step integrates our sort-based distributed de-duplication algorithm. This replaces the slow centralized CPU reduction with a high-bandwidth parallel exchange.
For the communication-heavy Cr$_2$ system, the de-duplication time decreases from 72.6s to 1.8s (\textbf{40$\times$ speedup}).
For N$_2$, it drops from 37.0s to 1.7s (\textbf{22$\times$ speedup}).

\textbf{Conclusion: Bottleneck Shift and Total Gain}
Combining these optimizations, QiankunNet-cuSCI achieves total end-to-end speedup of \textbf{1.88$\times$} for Cr$_2$ (239.1s to 127.0s) and \textbf{2.32$\times$} for N$_2$ (77.9s to 33.5s). More importantly, the system behavior has fundamentally shifted. In the optimized workflow (the rightmost bars), the AI inference stage (blue) once again dominates the runtime—accounting for \textbf{81\%} of the total time for Cr$_2$ and \textbf{76\%} for N$_2$. This confirms that QiankunNet-cuSCI successfully eliminates the scalability walls of traditional SCI components, allowing the system to fully utilize the massive tensor compute power of modern GPU clusters.

\subsection{Evaluation of Distributed De-duplication}\label{sec:eva_dis_de}

\begin{table}[t]
    \caption{Load balance and throughput of our global de-duplication algorithm across various chemical systems and GPU counts. The consistently low Max/Min Ratios and Coefficients of Variation (CV) demonstrate the robustness of our data-aware regular sampling method. Throughput is measured in M items/sec.
    }
    \centering
    \scriptsize 
    \resizebox{\linewidth}{!}{
    \renewcommand{\arraystretch}{1}
        \begin{tabular}{ccccc}
            \toprule[0.5mm]
            \textbf{System} & \textbf{GPUs} & \textbf{Max/Min} & \multirow{2}{*}{\textbf{CV}} & \textbf{Throughput} \\ 
            \textbf{(Unique Items)} & \textbf{Count} & \textbf{Ratio} & & \textbf{(M items/sec)} \\
            \midrule[0.4mm]
            \multirow{2}*{C$_{2}$H$_{4}$O (430.8 M)} & 32 & 1.10$\times$ & 0.021 & 1641.3 \\
             & 64 & 1.25$\times$ & 0.027 & 1391.1 \\
            \midrule
            \multirow{2}*{N$_{2}$ (320.4 M)} & 32 & 1.12$\times$ & 0.027 & 1175.4 \\
             & 64 & 1.20$\times$ & 0.025 & 1163.6 \\
            \midrule
            \multirow{2}*{C$_{2}$ (257.4 M)} & 32 & 1.11$\times$ & 0.024 & 985.9 \\
             & 64 & 1.02$\times$ & 0.010 & 1292.0 \\
            \midrule
            \multirow{2}*{H$_{2}$O (161.9 M)} & 32 & 1.11$\times$ & 0.022 & 693.8 \\
             & 64 & 1.03$\times$ & 0.012 & 695.5 \\
            \midrule
            \multirow{2}*{Cr$_{2}$ (966.5 M)} & 32 & 1.01$\times$ & 0.011 & 237.5 \\
             & 64 & 1.01$\times$ & 0.012 & 235.7 \\
            \bottomrule[0.5mm]
        \end{tabular}
    }
    \label{tab:load_balance_robustness}
\end{table}

A scalable, load-balanced global de-duplication algorithm is critical for eliminating inter-node redundancy and ensuring high parallel efficiency. We evaluate the performance of our sort-based algorithm, which leverages a \textit{regular sampling} strategy, across a variety of chemical systems and cluster sizes (32 and 64 NVIDIA A100 GPUs). We measure two key metrics to validate its design: (1) the load balance, quantified by the ratio of the maximum to minimum items per GPU after data exchange and the coefficient of variation (CV). (2) the input throughput, defined as the total number of redundant items processed per second. The results, summarized in Table \ref{tab:load_balance_robustness}, demonstrate the exceptional robustness and efficiency of our approach, mitigating data skew across diverse workloads.

First, our algorithm consistently achieves near-perfect load balance. Across all five chemically diverse systems—ranging from small molecules to the challenging 84-qubit Chromium Dimer (Cr$_{2}$) system with nearly 1 billion unique configurations—the Max/Min Ratio remains consistently low. Notably, for the largest Cr$_{2}$ system, the ratio approaches an ideal 1.01x with a CV of approximately 0.01 at both 32 and 64 GPU scales. This empirically proves that our regular sampling method effectively captures the global data distribution without requiring complex statistical estimation. Such deterministic load balancing is fundamental for the predictable scalability of our framework, effectively preventing straggler nodes that would otherwise bottleneck the computation.

Second, the algorithm maintains stable throughput across varying scales. For medium-sized systems like C$_{2}$H$_{4}$O, the algorithm sustains a throughput of over 1.3 billion items per second. Even for the computationally heavier Cr$_{2}$ system, it maintains consistent performance as the GPU count doubles. This efficiency stems from our design choices: the lightweight local sorting and sampling, the minimal overhead of gathering small sample sets (negligible compared to the data volume), and the highly optimized, single \texttt{MPI\_AlltoAllv} data exchange phase. 

In summary, this analysis validates our distributed de-duplication algorithm successfully meets its core design goals. It provides a robustly load-balanced and scalable solution that effectively removes the critical bottleneck of inter-node redundancy, even for large-scale systems with heavy workloads.

\subsection{Scalability Studies} \label{sec:eva_scal}

\begin{figure}[t]
    \centering
    \includegraphics[width=1\columnwidth]{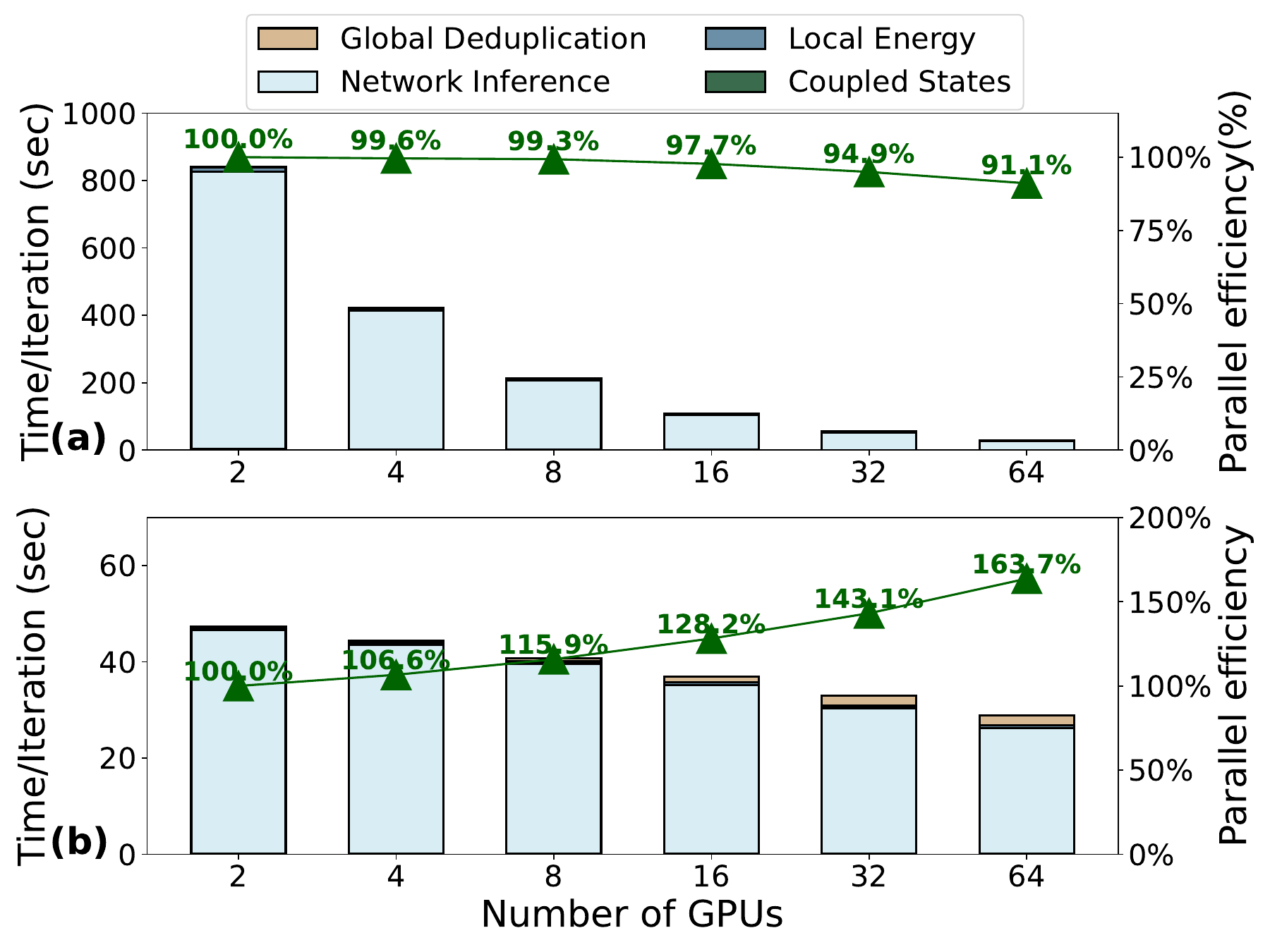}
    \Description{A two-panel chart illustrating the scalability of the N2 system across varying numbers of GPUs, from 2 to 64. 

    The top panel, (a), shows Strong scaling. A stacked bar chart represents the time per iteration in seconds, which halves consistently as the GPU count doubles, dropping from over 800 seconds to near zero. A superimposed line graph with triangular markers represents parallel efficiency, which stays remarkably flat and high, starting at 100 percent and maintaining 91.1 percent even at 64 GPUs. The dominant component of the bars is "Network Inference".

    The bottom panel, (b), shows Weak scaling. The stacked bar chart represents time per iteration, which surprisingly decreases slightly rather than staying constant as the GPU count increases. The superimposed parallel efficiency line graph climbs steadily upwards, starting at 100 percent and reaching an impressive 163.7 percent at 64 GPUs. 

    In both panels, the bars are overwhelmingly composed of the "Network Inference" category, with other categories like "Global Deduplication" forming barely visible slivers at the top of the bars.}
    \caption{Scalability of N$_2$: (a) Strong scaling (SCI space = 256,000), (b) Weak scaling (SCI space = 4000/GPU).}
    \label{fig:scaling}
\end{figure}

We conducted comprehensive scalability benchmarks using the N$_2$ system on a cluster of up to 64 NVIDIA A100 GPUs. 

\textbf{Strong Scaling.} For the strong scaling evaluation, we used a fixed, large-scale problem: the N2 molecule in a cc-pVDZ basis set with a SCI space of 256,000 source configurations. We measured the time per iteration while increasing the number of GPUs from 2 to 64. We observe excellent strong scaling for QiankunNet-cuSCI in Figure \ref{fig:scaling}(a). The time per iteration decreases smoothly from 825 seconds on 2 GPUs to just 29 seconds on 64 GPUs. The parallel efficiency, plotted on the secondary y-axis, remains remarkably high, 
sustaining 91.06\% at 64 GPUs. This result is a direct validation of our system design. The lightweight nature of our global de-duplication algorithm (purple bar) and other overheads is minimal, preventing communication costs from overwhelming the parallel speedup. This allows the dominant, perfectly parallelizable Network Inference stage to scale almost linearly with the number of processors.

\begin{figure}[t]
    \centering
    \includegraphics[width=.9\columnwidth]{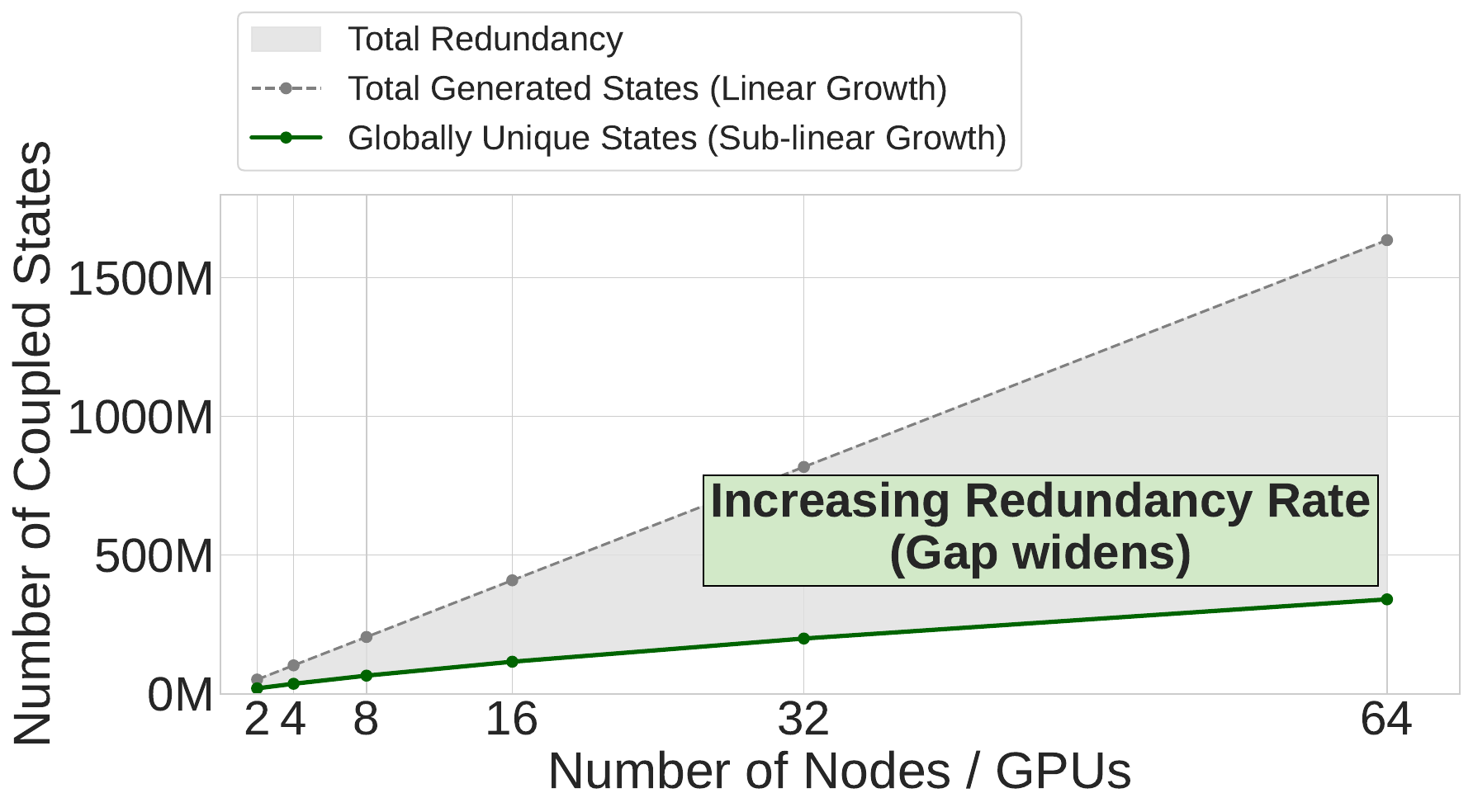}
    \Description{A line graph plotting the number of coupled states on the y-axis against the number of GPUs on the x-axis, ranging from 2 to 64. A top dashed line, representing Total Generated States, grows linearly and steeply. A bottom solid line, representing Globally Unique States, grows sub-linearly and much more slowly. The area between the two lines is shaded grey to represent Total Redundancy. This shaded gap widens significantly as the number of GPUs increases, visually highlighting the increasing redundancy rate, which is also emphasized by a text box pointing to the widening gap.}
    \caption{Non-linear growth of unique configurations in weak scaling test.}
    \label{fig:weak-scaling}
\end{figure}

\textbf{Weak Scaling.} For the weak scaling test, we defined the workload per GPU to be 4,000 source configurations and scaled the total problem size proportionally with the number of GPUs. The results, presented in Figure \ref{fig:scaling}(b), reveal a significant and important phenomenon: super-linear speedup. The time per iteration does not remain constant but decreases as the scale increases. This leads to an impressive parallel efficiency reaching 163.70\% on 64 GPUs. This counter-intuitive result is not primarily due to hardware artifacts like cache effects, but is a direct consequence of the interplay between the SCI problem's intrinsic redundancy and our highly efficient global de-duplication algorithm. To illustrate this, Figure \ref{fig:weak-scaling} plots the growth of the total generated coupled configurations versus the globally unique configurations during the weak scaling experiment. While the total number of generated configurations (the input workload) grows linearly with the number of GPUs, the number of globally unique configurations (the effective computational workload) exhibits a distinct sub-linear growth.

The widening gap between these two curves demonstrates that the global redundancy rate increases dramatically at larger scales. For example, at 64 GPUs, over 66\% of the generated configurations are redundant. Because our global de-duplication algorithm efficiently eliminates all of this redundancy, the effective computational workload that each GPU must process in the expensive Network Inference stage actually decreases as we add more nodes. This reduction in the true workload per node directly leads to a shorter execution time per iteration and, therefore, a measured parallel efficiency exceeding 100\%. This finding powerfully underscores that an efficient global de-duplication strategy is not merely an optimization for SCI simulations; it is a critical component that enables
more favorable scaling dynamics at large scales.

\subsection{Evaluation of Memory Management } \label{sec:eva_mem}

The ultimate goal of our framework is to break the node-level memory wall, enabling simulations of systems that are too large to fit into a single GPU's memory. To validate the effectiveness of our three-stage execution model, we selected four representative chemical systems and configured them with 50,000 configurations in SCI space to create a memory-intensive workload.

Theoretical Peak (W/O Optimization): This represents the peak GPU memory required if all intermediate data for a full iteration (e.g., all coupled configurations, reverse indices, and $Psi$ values) had to reside simultaneously in GPU memory. This value is calculated based on the data structures and serves as the baseline for a conventional, non-optimized implementation.

QiankunNet-cuSCI Measured Peak (W/ Optimization): This is the actual, maximum GPU memory allocated during a full iteration of our framework. We measure this using torch api. Since our custom CUDA kernels are integrated as PyTorch extensions and other components (like CuPy) are configured to use PyTorch caching allocator, this provides a reliable, holistic measurement of the peak memory usage for the entire workflow. All experiments were conducted on an NVIDIA A100 GPU with a 40 GB GPU memory capacity limit.

\begin{figure}[t]
    \centering
    \includegraphics[width=.9\columnwidth]{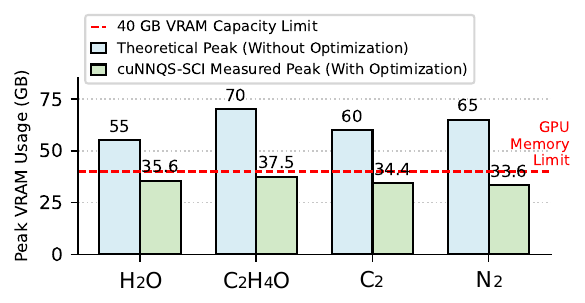}
    \Description{A grouped bar chart comparing Peak VRAM Usage for four systems: H2O, C2H4O, C2, and N2. A red dashed horizontal line prominently marks the 40 GB GPU memory limit. For every system, the "Theoretical Peak" bar stands significantly above this 40 GB limit line, ranging from 55 to 70 GB. Right next to them, the "QiankunNet-cuSCI Measured Peak" bars are all safely below the red line, ranging from 33.6 to 37.5 GB. This visually proves that the optimization successfully keeps the memory footprint within the hardware constraints.}
    \caption{
    Comparison of theoretical and QiankunNet-cuSCI measured peak device memory usage, showing that QiankunNet-cuSCI keeps usage within hardware limits.
    }
    \label{fig:memory}
\end{figure}

Figure \ref{fig:memory} powerfully demonstrates the efficacy of our design. For all four tested systems, the Theoretical Peak GPU memory (red line) significantly exceeds the 40 GB physical capacity limit of the GPU. For instance, the C$_{2}$H$_{4}$O and N$_{2}$ systems would theoretically require 70 GB and 65 GB of GPU memory, respectively. This clearly indicates that these scientifically important calculations would be entirely infeasible with a conventional approach, leading to an immediate out-of-memory error. In contrast, the QiankunNet-cuSCI Measured Peak (blue bars) stays comfortably below the hardware limit in every case. Our three-stage pipeline, which intelligently stages data between host memory and device memory, successfully reduces the peak memory footprint to a manageable level. For  C$_{2}$H$_{4}$O , the peak usage was reduced by 46.4\% to just 37.5 GB. For N$_{2}$, the reduction was 48.3\%, bringing the peak down to 33.6 GB.

Notably, the measured peaks are intentionally close to the 40 GB limit. This is not a sign of inefficiency, but rather a feature of our design: we aim to maximize the utilization of available GPU memory by creating the largest possible data batches for each stage. This ensures high computational memory usage while guaranteeing the process remains within physical memory bounds.

\section{Related Works}

\textbf{Neural Network Quantum States (NNQS) Scaling.} Since the introduction of the RBM ansatz by Carleo and Troyer~\cite{carleo2017solving}, NNQS has evolved from small lattice prototypes to massive ab-initio simulations requiring leadership-class supercomputing. Early frameworks such as NetKet~\cite{carleo2019netket} primarily utilized MPI parallelism over Monte Carlo chains. With the rise of deep architectures like FermiNet~\cite{pfau2020ab} and PauliNet~\cite{hermann2020deep}, the focus shifted toward GPU acceleration. Recent works exemplify this transition to large-scale GPU computing. Vicentini et al.~\cite{vicentini2022netket} systematically integrated MPI-based distributed stochastic reconfiguration (SR) protocols into the NetKet 3 toolkit; Zhao et al.~\cite{Zhao2022AI4QM_SC22} employed batch autoregressive sampling, scaling to systems of 76 qubits. However, these approaches predominantly rely on variational Monte Carlo (VMC) sampling, which suffers from stochastic noise rooted in the fermionic sign problem~\cite{sandvik2010computational, choo2020fermionic, OGorman2022Intractability}.

In contrast, QiankunNet-cuSCI adopts a deterministic SCI strategy. Unlike sampling-based methods that heuristically infer the wavefunction, our approach systematically constructs the Hilbert space. While this avoids sampling noise, it introduces irregular memory access patterns and dynamic workload imbalance that strictly sampling-based frameworks (e.g.,~\cite{qknet2023, ma2024quantum}) do not address.

\noindent\textbf{GPU-Accelerated Selected CI. }Selected CI (SCI) methods, originating from the CIPSI algorithm~\cite{huron1973iterative} and including modern variants such as Heat-bath CI (HCI) and Adaptive Sampling CI (ASCI)~\cite{Sharma2017SHCI,tubman2020modern}, reduce computational cost by filtering the configuration space. Traditional CI implementations, like NWChem~\cite{valiev2010nwchem}, are largely CPU-bound due to the complex logic and irregular memory access patterns of Slater-Condon evaluations. Early GPU acceleration efforts offloaded only numerically intensive kernels~\cite{ufimtsev2009quantum}, leaving configurations generation and management on the host, which made PCIe communication and CPU throughput scalability bottlenecks. QiankunNet-SCI~\cite{Kan2025NNQS} combines NNQS inference with SCI but retains a heterogeneous execution model that limits scalability to large candidate spaces.

The fundamental challenge of GPU-based SCI lies in the highly irregular configuration expansion, leading to thread divergence, uncoalesced memory access, and severe workload imbalance, similar to GPU graph traversal and sparse matrix expansion. While optimizations for specific Hamiltonians exist~\cite{whitfield2013computational}, general-purpose GPU kernels for on-the-fly Slater–Condon evaluation remain scarce. QiankunNet-cuSCI addresses these challenges by moving the entire SCI workflow—generation, coupling, deduplication, and inference—onto the GPU. It also redesigns data layouts for irregular workloads, drawing inspiration from high-performance graph analytics frameworks~\cite{wang2016gunrock} and the GAP Benchmark Suite~\cite{beamer2015gap}.

\noindent\textbf{Memory-Centric System Design.} Memory capacity is the primary constraint for high-accuracy quantum simulations on GPUs. Standard approaches, particularly in tensor-network states like DMRG~\cite{schollwock2011density} or in full CI, are fundamentally limited by memory capacity and run out of memory (OOM) for large systems. In deep learning, techniques to overcome this include pipeline parallelism in GPipe~\cite{huang2019gpipe} or memory optimization with CPU-offloading in ZeRO~\cite{ren2021zero}. However, the random access nature of checking if a configuration exists in the selected space makes standard swapping inefficient for SCI. Recent system-level optimizations for NNQS, such as those discussed in~\cite{zhao2023scalable} and~\cite{Sharir2020DeepAutoregressive}, focus on optimizing network weights memory. QiankunNet-cuSCI shifts the focus to the state memory. By implementing a memory-centric execution model that dynamically streams mini-batches of configurations, we treat GPU memory as a cache for the active Hilbert space. This allows QiankunNet-cuSCI to simulate system sizes that exceed the physical HBM capacity of a single device, extending the reach of GPU-accelerated SCI beyond what was possible with purely in-core approaches like NNBF~\cite{Liu2024NNBF}.

\section{Conclusion and Future Work}
\label{sec:conclusion}
QiankunNet-cuSCI overcomes the long-standing scalability bottleneck in high-precision quantum chemistry methods like SCI by addressing the node-level memory wall. Our memory-first design incorporates: (1) memory-optimized CUDA kernels for core computation, (2) a load-balanced global de-duplication algorithm to eliminate redundancy, and (3) a three-stage out-of-core execution model that exceeds single-GPU memory limits. Evaluation demonstrates over 90\% parallel efficiency on 64 GPUs and substantial speedups over SOTA heterogeneous solutions.

Our future work will focus on further enhancing the scalability of QiankunNet-cuSCI by refining its core architectural components. We plan to investigate more sophisticated memory management strategies, to further reduce the memory pressure from large, intermediate data structures. Concurrently, we will explore advanced network communication methods, to minimize latency and improve parallel efficiency on ultra-large-scale clusters. These combined efforts will continue to push the boundaries of what is possible, enabling the application of QiankunNet-cuSCI to even more complex and scientifically significant large-scale systems.

\begin{acks}
This work was supported by the National Natural Science Foundation of China (Grant Nos. 62032023 and T2125013), the Innovation Funding of ICT, CAS (Grant No. E461050), and the National Key Research and Development Program of China (Grant No. 2025YFB3003702). The AI-driven experiments, simulations and model training were performed on the robotic AI-Scientist platform of Chinese Academy of Sciences.
\end{acks}

\bibliographystyle{ACM-Reference-Format}
\bibliography{references}

\end{document}